\pdfoutput=1
\documentclass{article}
\usepackage{jheppub}

\usepackage{packages}
\usepackage{notations}
\usepackage{amssymb}
\usepackage{extarrows}

\def\C{\mathcal{N}}
\def\UHP{\mathbb{H}}
\def\Cp{f_{0}}
\def\F{\alpha}
\def\zb{\bar{z}}
\def\r{x}
\def\sigmam{\sigma}
\def\sigmap{\varphi}
\def\g{r}

\DeclareMathOperator*{\SumInt}{%
\mathchoice%
  {\ooalign{$\displaystyle\sum$\cr\hidewidth$\displaystyle\int$\hidewidth\cr}}
  {\ooalign{\raisebox{.14\height}{\scalebox{.7}{$\textstyle\sum$}}\cr\hidewidth$\textstyle\int$\hidewidth\cr}}
  {\ooalign{\raisebox{.2\height}{\scalebox{.6}{$\scriptstyle\sum$}}\cr$\scriptstyle\int$\cr}}
  {\ooalign{\raisebox{.2\height}{\scalebox{.6}{$\scriptstyle\sum$}}\cr$\scriptstyle\int$\cr}}
}

\tikzset{
    partial ellipse/.style args={#1:#2:#3}{
        insert path={+ (#1:#3) arc (#1:#2:#3)}
    }
}

\tikzset{middlearrow/.style={
        decoration={markings,
            mark= at position 0.55 with {\arrow[thick]{#1}} ,
        },
        postaction={decorate}
    }
}

\title{\boldmath Fishnet four-point integrals: integrable representations and thermodynamic limits}
\author{Benjamin Basso$^a$, Lance J.~Dixon$^b$, David A.~Kosower$^c$, Alexandre Krajenbrink$^{d}$, De-liang Zhong$^e$}

\affiliation{$^a$Laboratoire de Physique de l'Ecole Normale Sup\'erieure, ENS, Universit\'e PSL, CNRS, Sorbonne Universit\'e, Universit\'e de Paris,
F-75005 Paris, France\\
$^b$SLAC National Accelerator Laboratory,
Stanford University, Stanford, CA 94309, USA\\
$^c$Institut de Physique Th\'eorique, CEA, CNRS, Universit\'e Paris-Saclay, F-91191 Gif-sur-Yvette cedex, France\\
$^d$SISSA and INFN, via Bonomea 265, 34136 Trieste, Italy\\
$^e$School of Physics and Astronomy, Tel Aviv University, Ramat Aviv 69978, Israel}

\abstract{We consider four-point integrals arising in the planar limit of the conformal ``fishnet''  theory in four dimensions. They define a two-parameter family of higher-loop Feynman integrals, which extend the series of ladder integrals and were argued, based on integrability and analyticity, to admit matrix-model-like integral and determinantal representations. In this paper, we prove the equivalence of all these representations using exact summation and integration techniques. We then analyze the large-order behaviour, corresponding to the thermodynamic limit of a large fishnet graph. The saddle-point equations are found to match known two-cut singular equations arising in matrix models, enabling us to obtain a concise parametric expression for the free-energy density in terms of complete elliptic integrals. Interestingly, the latter depends non-trivially on the fishnet aspect ratio and differs from a scaling formula due to Zamolodchikov for large periodic fishnets, suggesting a strong sensitivity to the boundary conditions. We also find an intriguing connection between the saddle-point equation and the equation describing the Frolov-Tseytlin spinning string in $AdS_{3}\times S^{1}$, in a generalized scaling combining the thermodynamic and short-distance limits.}

\preprint{ \begin{flushright} SLAC--PUB--17600 \end{flushright}}

\begin{document}
\maketitle

\section{Introduction}

Over the past few years, we have witnessed tremendous progress in our understanding of mathematical and physical properties of Feynman integrals, notably in the massless limit relevant for the description of high-energy processes. A great deal of this progress has centred around the most symmetrical and solvable conformal field theory in four dimensions, planar $\mathcal{N}=4$ super-Yang-Mills (SYM) theory, where dualities, conformal symmetry and integrability paved the way to the development of a variety of new methods for efficient higher-loop calculations.

It has also been realized that these methods, including integrability, are not restricted to supersymmetric planar field theories. They extend to a large class of interacting non-supersymmetric planar theories, the fishnet conformal field theories \cite{Gurdogan:2015csr,Caetano:2016ydc,Mamroud:2017uyz,Gromov:2017cja,Grabner:2017pgm,Kazakov:2018qbr,Derkachov:2018rot,Kazakov:2018gcy,Pittelli:2019ceq}, which admit formulations in any spacetime dimensions and may be realized in four dimensions by deforming $\mathcal{N}=4$ SYM, in ways that preserve conformal symmetry and integrability. This web of theories allows us to shed light on properties of a broad family of higher-loop planar integrals, with regular iterative structures, the so-called fishnet graphs. These diagrams are endowed with integrable structures, such as Yangian symmetry~\cite{Chicherin:2017frs,Chicherin:2017cns,Loebbert:2020hxk,Loebbert:2020tje}, which put severe constraints on their analytic expressions and facilitate their determination. 

\begin{figure}[t]
\begin{center}
\includegraphics[scale=0.5]{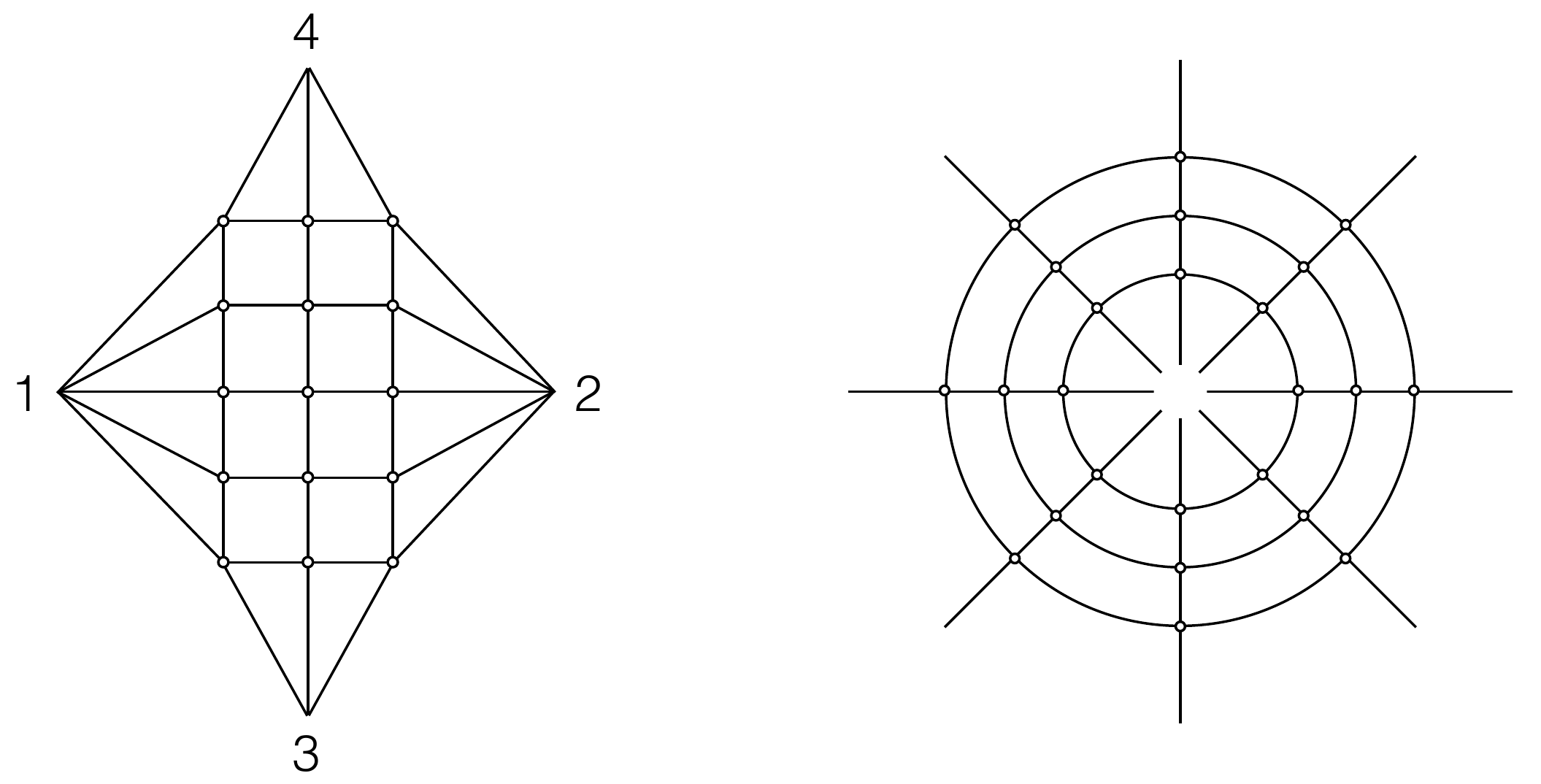}
\end{center}
\caption{Specimens of planar fishnet graphs in position space with different boundary conditions. Bulk points stand for quartic integration points, $\int d^{4}x/\pi^{2}$, while edges are massless propagators, $1/(x-y)^2$. For the fishnet four-point functions considered in this paper, the $m$ vertical and $n$ horizontal external legs end on four fixed spacetime points $x_{1,2,3,4}$, as shown in the left panel ($m$ lines end on $x_{3,4}$; $n$ lines end on $x_{1,2}$). In the right panel, we show an $m$-by-$n$ fishnet graph with periodic boundary conditions in the angular direction. Such graphs are expected to scale as $\sim g_c^{-2mn}$ when $m, n \rightarrow \infty$, with $g_c$ a numerical constant \cite{Zamolodchikov:1980mb}. The four-point function considered in this paper (left panel) scales differently, with a free energy per vertex that depends on the aspect ratio $k = n/m$.} \label{BCs}
\end{figure}

In this paper, we will examine a simple class of planar fishnet integrals in four dimensions, associated with regular square lattices attached to four generic spacetime points, as shown in the left panel of figure~\ref{BCs}. They generalize the well-known ladder conformal integrals, calculated many years ago~\cite{Usyukina:1993ch}. They are subject to stringent analytic properties and were argued to admit integrability-based integral and determinantal representations~\cite{Basso:2017jwq}, some of which have been proven recently \cite{Derkachov:2019tzo,Derkachov:2020zvv,Derkachov:2021rrf}. The first goal of this paper will be to place the proposals in ref.~\cite{Basso:2017jwq} on a firmer footing by proving their mathematical equivalence. To be precise, we will show the equivalence of the so-called Berenstein-Maldacena-Nastase (BMN) \cite{Berenstein:2002jq} and flux-tube (FT) integral representations with the determinant of a Hankel matrix of ladder integrals, using exact summation and integration techniques.

The integrability of fishnet graphs was noticed many years ago by Zamolodchikov \cite{Zamolodchikov:1980mb}, who considered them as an example of exactly solvable lattice models and calculated their leading behaviour in the thermodynamic limit for periodic boundary conditions. Taking this ``continuum'' limit was partly motivated by viewing fishnet graphs as approximating the worldsheet of a string, see e.g.~ref.~\cite{Sakita:1970ep}. Recent studies have given support to this picture, in a more holographic setting, connecting the conformal fishnet integrals to quantum mechanical systems in Anti-de Sitter space (AdS) \cite{Basso:2018agi,Gromov:2019aku,Gromov:2019bsj,Gromov:2019jfh,Basso:2019xay}. In particular, in the continuum limit, evidence was found that tubular fishnet diagrams, like the one shown in the right panel of figure~\ref{BCs}, admit an effective description in terms of a two-dimensional (2d) non-linear sigma model in AdS.

Motivated by this consideration, in the second half of this paper, we study the thermodynamic limit of the four-point fishnet integrals. We do so by taking the dimensions of the four operators, or alternatively the lengths of each side of a rectangular fishnet, to be large, holding the aspect ratio (the ratio of side lengths) fixed. In this limit, the integrability-based integrals can be evaluated in a saddle-point approximation, and the saddle-point equations coincide with solvable problems encountered in the study of matrix models and integrable systems. This connection will allow us to obtain a closed parametric expression for the bulk free energy in terms of complete elliptic integrals. Interestingly, the free energy is independent of the cross ratios parametrizing the positions of the boundary operators. However, we will find that it depends in a complicated manner on the rectangular aspect ratio.  For any value of the aspect ratio, it differs from Zamolodchikov's result for periodic graphs, indicating a strong dependence of the thermodynamic limit on the boundary conditions. We shall also study a more general scaling regime, which combines the short-distance and thermodynamic limits; this limit reveals an intriguing connection with the equation describing a classical spinning string in $AdS_{3}\times S^{1}$, studied by Frolov and Tseytlin \cite{Frolov:2002av}.

This paper is structured as follows. In section \ref{sec:representations}, we analyze the integrability-based representations for the fishnet integrals and prove their equivalence with the determinant of ladder integrals. Our analysis relies on exact methods which may be applicable to more general fishnet graphs. We proceed with the analysis of the thermodynamic limit in section \ref{sec:thermodynamic}. Applying the saddle-point method to the matrix-model integrals gives us first order differential equations for the free-energy density, which can be solved simultaneously in terms of complete elliptic functions. We compare the obtained expression with a direct numerical evaluation of the determinant and study particular limits analytically. We conclude with a brief discussion of the dependence of the free energy on the boundary conditions. In section \ref{sec:spinning} we consider a more general scaling which combines the thermodynamic and short-distance limits and unveils a connection with the singular equation describing a classical spinning string in $AdS_{3}\times S^{1}$. Section \ref{sec:conclusion} contains concluding remarks. 

\section{Integral representations}\label{sec:representations}

The fabric of the fishnet graphs -- the conformal fishnet theory -- is a theory of two $N_{c}\times N_{c}$ matrices of complex scalar fields $\phi_{1,2}$ and a quartic interaction, with (Euclidean) Lagrangian density \cite{Gurdogan:2015csr,Caetano:2016ydc,Grabner:2017pgm}
\beq
\mathcal{L} = \textrm{Tr}\, \big\{ \partial_{\mu}\phi_{i}\partial_{\mu}\phi^{\dagger}_{i} -(4\pi g)^{2} \phi_{1}\phi_{2}\phi_{1}^{\dagger}\phi_{2}^{\dagger} \big\}\, ,
\eeq
with the trace running over the ``colour'' indices and with $g^2$ a marginal coupling constant, which is kept fixed in the large $N_{c}$ limit \cite{Grabner:2017pgm,Sieg:2016vap}.%
\footnote{A complete description of the planar limit entails introducing and fine-tuning double-trace couplings \cite{Grabner:2017pgm,Sieg:2016vap}; they only affect correlators of short-length single-trace operators and will not play any role here.}

\begin{figure}[t]
\begin{center}
\includegraphics[scale=0.45]{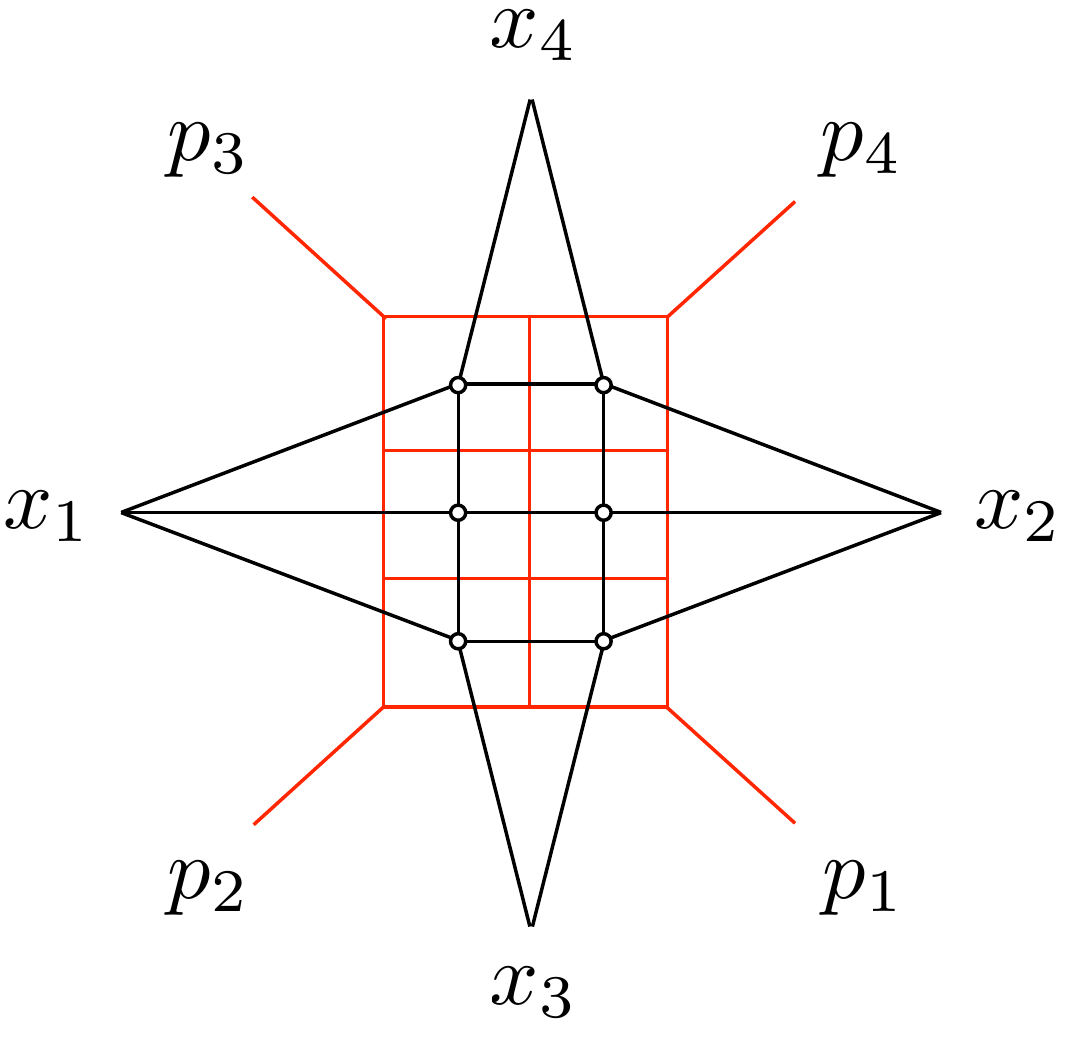}
\end{center}
\caption{A transformation maps the space-time integral into a momentum space amplitude, shown in red lines, with external massive momenta $p_{1} = x_{23}, p_{2} = x_{31}, p_{3}=x_{14}$ and $p_{4} =x_{42}$, and with massless internal lines. The latter amplitude is subject to important analyticity requirements such as the Steinmann relations.} \label{fig:fishnets}
\end{figure}

The fishnet four-point function of interest is defined as the leading colour-ordered contribution to the correlator
\begin{equation}
 G_{m,n}(\{x_{i}\}) = \langle \textrm{Tr} \{ \phi_2^n(x_1) \phi_1^m(x_3) \phi_2^{\dagger n}(x_2) \phi_1^{\dagger m} (x_4) \} \rangle\, ,
\end{equation}
with the trace embracing all the fields. It receives a single contribution at large $N_c$, which is given by the $mn$-loop Feynman diagram shown in figure \ref{fig:fishnets}.

The graph is both ultraviolet (UV) and infrared (IR) finite, for generic $x_{1,2,3,4}$. Dropping a colour factor and stripping off overall weights, one has
\beq
G_{m, n}(\{x_{i}\}) = \frac{g^{2mn}}{(x_{12}^{2})^n (x_{34}^{2})^m} \times \Phi_{m, n}(u, v)\, ,
\eeq
with  $x_{ij}^2 \equiv (x_{i}-x_{j})^2$, where $\Phi_{m, n}$ is a coupling-independent function of the two conformal cross ratios
\beq
u = \frac{x_{14}^2 x_{23}^2}{x_{12}^2 x_{34}^2} \equiv \frac{z \bar{z}}{(1-z)(1-\bar{z})}, \qquad v = \frac{x_{13}^2 x_{24}^2}{x_{12}^2 x_{34}^2} \equiv \frac{u}{z \bar{z}}\, ,
\eeq
parametrized here in terms of two numbers $z, \bar{z}$. The latter are complex conjugates in Euclidean spacetime signature, and are treated as independent real numbers in the Minkowskian case.

A representation for $\Phi_{m, n}$ was proposed in ref.~\cite{Basso:2017jwq} using integrability and analyticity. The proposal is that for $n\geqslant m$
\beq\label{eq:Phi-I}
\Phi_{m, n} = \left[ \frac{(1-z)(1-\bar{z})}{z-\bar{z}} \right]^m I_{m, n}(z, \bar{z})\, ,
\eeq
where $I_{m, n}(z, \bar{z})$ is a pure function of weight $m\times n$ given by the determinant of an $m\times m$ Hankel matrix of ladder functions. It reads
\beq\label{eq:I-det}
I_{m, n} = \frac{1}{\C} \det_{1\leqslant i, j \leqslant m} (M_{i+j+n-m-1})\, ,
\eeq
with the matrix element
\beq \label{eq:valueM}
M_{p} = p! (p-1) !  L_{p}(z, \bar{z})\, ,
\eeq
and normalization factor
\beq\label{eq:calN}
\C = \prod_{k=0}^{2m-1}(n-m+k)!\, .
\eeq
The particular case $m=1$ coincides with the well-known ladder integral $L_{p}(z, \bar{z})$, defined for a number of rungs $p \geqslant 0$ by \cite{Usyukina:1993ch,Broadhurst:2010ds}
\be\label{ladders}
L_p(z,\bar{z}) = \begin{dcases}
\sum_{j=p}^{2p} \frac{j![-\log{(z \bar{z})}]^{2p-j}}{p!(j-p)!(2p-j)!} [\text{Li}_j(z) - \text{Li}_j(\bar{z})] & \text{if}\ p \geqslant 1, \\
\frac{z-\bar{z}}{(1-z)(1-\bar{z})} & \text{if}\ p=0,
\end{dcases}
\ee
with $\textrm{Li}_{j}(z)$ the polylogarithm of weight $j$,
\beq\label{polylog}
\text{Li}_j(z) = \frac{z}{(j-1)!} \int\limits_{0}^{\infty} \rmd r  \frac{r^{j-1}}{e^{r}-z} = \sum_{k=1}^\infty \frac{z^k}{k^j}\, ,
\eeq
where the Taylor series converges for $|z|<1$.

The determinant formula~\eqref{eq:I-det}, when interpreted as the result for a scattering amplitude rather than a correlation function, nicely satisfies the Steinmann relations~\cite{Steinmann,Steinmann2}, which forbid sequential discontinuities in overlapping channels.  In figure \ref{fig:fishnets}, the red lines indicate the scattering interpretation, and the two channels are $(p_1+p_2)^2 = x_{12}^2$ and $(p_2+p_3)^2=x_{34}^2$.  The Steinmann relations imply that
\be\label{Steinmannx}
{\rm Disc}_{x_{12}^2} {\rm Disc}_{x_{34}^2} I_{m,n} = 0,
\ee
in the region where $x_{12}^2$ and $x_{34}^2$ both vanish, which is the neighborhood of $z,\zb\rightarrow1$.  Conformal invariance then translates eq.~\eqref{Steinmannx} into the vanishing of the double discontinuity in $(1-z)(1-\zb)$:
\be\label{Steinmannz}
{\rm Disc}_{1} {\rm Disc}_{1} I_{m,n} = 0.
\ee
(For example, one can wrap $z$ around 1, and keep $\zb$ fixed, or wrap
both in the same direction.)
The single discontinuity of the ladder integral~\eqref{ladders}
is~\cite{Basso:2017jwq,Bourjaily:2020wvq}
\be
{\rm Disc}_1 L_p =
2\pi \I  \frac{(-1)^p}{p!(p-1)!} \, \log(z/\zb) \, ( \log z \log\zb )^{p-1} \, ,
\label{Ldisc}
\ee
with $\I = \sqrt{-1}$. Since the single discontinuity
contains only $\log z$ and $\log \zb$, the double discontinuity of
any ladder integral vanishes, $[{\rm Disc}_1]^2 L_p  = 0$.
Furthermore, the discontinuities of any two columns (or two rows) of the
matrix $M_p$ defined in \eqref{eq:valueM} are proportional, because
\be
{\rm Disc}_1 M_{p+1} = (- \log z \log \zb ) \, {\rm Disc}_1 M_{p} \,,
\label{DiscM}
\ee
which guarantees the vanishing of the double discontinuity of the determinant~\eqref{eq:I-det}~\cite{Basso:2017jwq}. (A more general solution to the Steinmann relations is given by the minors of a semi-infinite matrix of ladders, as found in the study of large-charge correlators in $\mathcal{N}=4$ SYM~\cite{Coronado:2018cxj,Coronado:2018ypq}.)

\begin{figure}[t]
\begin{center}
\includegraphics[width=0.50\textwidth]{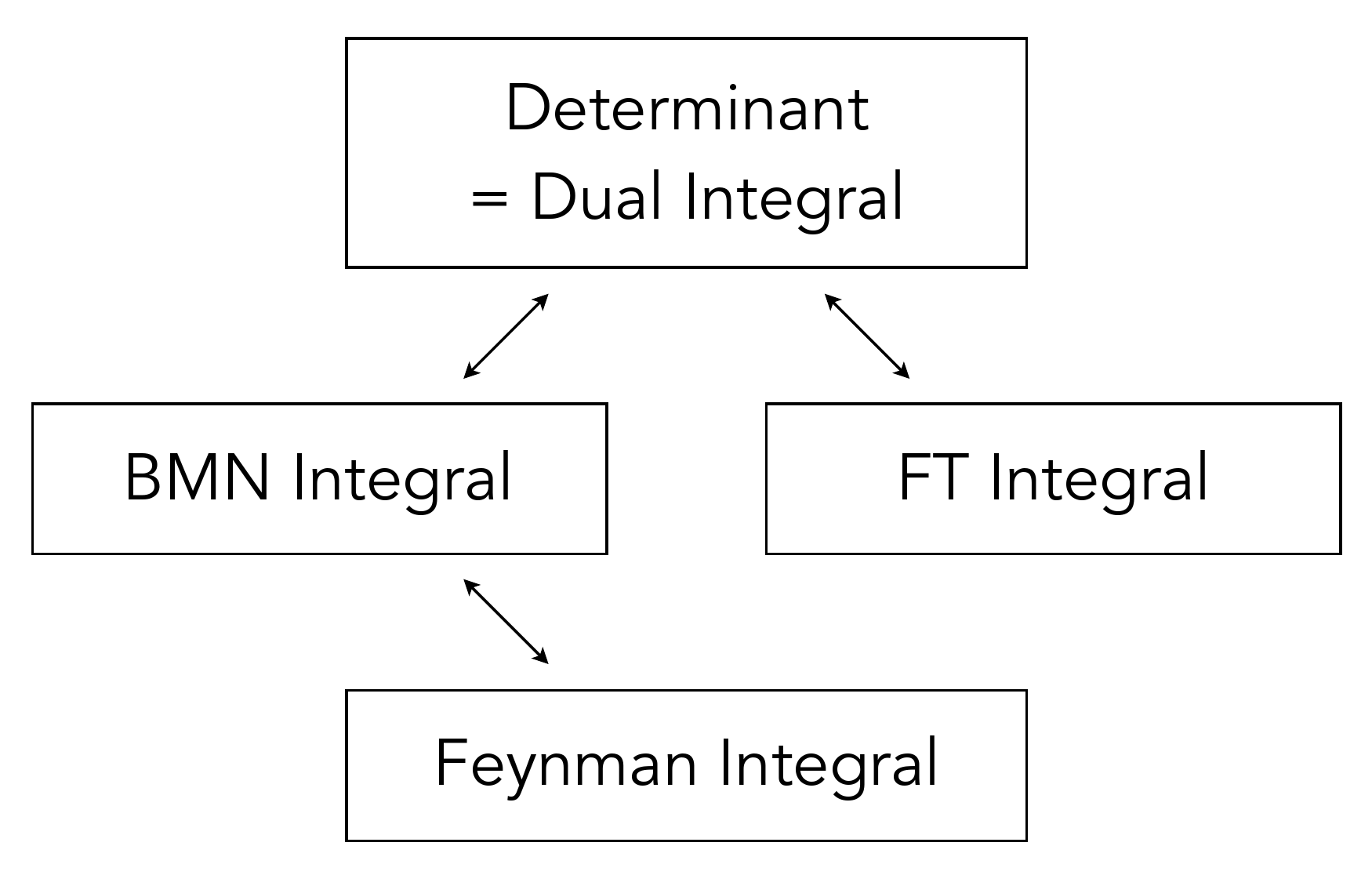}
\end{center}
\caption{Representations of the fishnet four-point integral. The Feynman integral was recently proved to be equivalent to the BMN integral~(\ref{eq:I_BMN}) in refs.~\cite{Derkachov:2019tzo,Derkachov:2020zvv}. In this paper we concentrate on the pyramid at the top and derive the determinant formula from the BMN and FT integrability-based representations.}\label{fig:representations}
\end{figure}

As alluded to before, along with the determinant~\eqref{eq:I-det}, two matrix-model-like integrals, the BMN integral and FT integral, were given in ref.~\cite{Basso:2017jwq}. They originate from two different conjectural $\mathcal{N}=4$ SYM integrability-based tiling constructions, the ``hexagonalization'' \cite{Fleury:2016ykk,Eden:2016xvg,Fleury:2017eph,Basso:2015zoa,Basso:2018cvy} and the ``pentagon OPE'' \cite{Basso:2013vsa,Alday:2007mf,Alday:2010ku}. The former has been receiving a lot of attention recently, as it is closely related to the method of separation of variables for conformal integrable spin chains, which enabled the calculation of fishnet correlators in the 2d fishnet conformal field theories~\cite{Derkachov:2018rot}. This method was recently extended to four dimensions in refs.~\cite{Derkachov:2019tzo,Basso:2019xay,Derkachov:2020zvv,Chicherin:2012yn}, allowing to put on firm ground the results obtained from the $\mathcal{N}=4$ SYM integrability-based tiling constructions. According to these analyses, the BMN representation discussed below is no longer a conjecture -- it is proved to be equivalent to the Feynman integral.
In this section we will close the remaining gaps to the determinant formula by proving its equivalence with both the BMN and FT integral representations, as schematized in figure \ref{fig:representations}.

\subsection{Dual integral}\label{sec:integralDet}
To begin, let us recast the determinant formula~\eqref{eq:Phi-I} into an integral form which will prove very useful later on. Introducing the parametrization
\beq\label{eq:z-mapping}
z =  -e^{\sigmam+\sigmap}\, , \qquad  \bar{z} = -e^{\sigmam-\sigmap}\, ,
\eeq
we will show in this subsection that
one may write the determinant in the form
\beq\label{eqn-detInt}
\Phi_{m, n} = d(z, \bar{z})^m  \times I^{{\rm Dual}}_{m, n}\, ,
\eeq
where $I^{{\rm Dual}}_{m, n}$ stands for the $m$-fold integral
\beq\label{eq:Idual}
I^{{\rm Dual}}_{m, n} = \frac{1}{2^m m! \mathcal{N}} \prod_{i=1}^{m}\int_{|\sigmam|}^{\infty} \frac{\rmd x_{i} x_{i}(x_{i}^2-\sigmam^2)^{n-m}}{\cosh{\tfrac{1}{2}(x_{i}+\sigmap)}\cosh{\tfrac{1}{2}(x_{i}-\sigmap)}} \prod_{i< j}^{m} (x_{i}^2-x_{j}^2)^2\, ,
\eeq
with $\C$ given in eq.~\eqref{eq:calN}, and with the kinematical prefactor
\beq\label{Gmn}
d(z, \bar{z}) = \frac{(1-z)(1-\bar{z})}{\sqrt{z\bar{z}}}\, .
\eeq
Note that this prefactor differs from the one in eq.~\eqref{eq:Phi-I}, such that
\be\label{eq:IdualIrelation}
I^{{\rm Dual}}_{m, n} = \left[\frac{\sqrt{z\bar{z}}}{z-\bar{z}}\right]^m I_{m, n}\, .
\ee

The representation~\eqref{eqn-detInt} may be seen as a reduction of the more general functional determinant formulae obtained in refs.~\cite{Kostov:2019stn,Kostov:2019auq,Belitsky:2019fan,Belitsky:2020qrm,Belitsky:2020qir,Kostov:2021omc} for large-charge correlators in $\mathcal{N}=4$ SYM theory, which resum infinite series of fishnet diagrams and generalizations thereof \cite{Coronado:2018ypq,Coronado:2018cxj}. In the following, we shall refer to eq.~\eqref{eq:Idual} as the ``dual integral'', as it relates in our set-up to Fourier-like transforms of the integrability-based integrals to be studied later on.

The dual representation given in eqs.~\eqref{eqn-detInt} and~(\ref{eq:Idual}) holds for $\sigmam \in \mathbb{R}$ and $\textrm{Im}\, \sigmap \in (-\pi, \pi)$, which includes both the Minkowskian and the Euclidean kinematics, corresponding respectively to purely real and purely imaginary values of $\sigmap$. It can be derived from the determinant~\eqref{eq:I-det} by first replacing the ladder functions by their integral representation \cite{Broadhurst:2010ds,Kostov:2019stn,Kostov:2019auq}
\beq\label{int-ladd}
L_p(z,\bar{z}) = \frac{(z-\bar{z})}{2\sqrt{z\bar{z}}\, p! (p-1)!} \int\limits_{|\sigmam|}^{\infty} \frac{\rmd x x(x^2-\sigmam^2)^{p-1}}{\cosh{\tfrac{1}{2}(x+\sigmap)}\cosh{\tfrac{1}{2}(x-\sigmap)}} \, .
\eeq
This identity follows directly from the definition~\eqref{polylog} of the polylogarithms. Namely, plugging eq.~\eqref{polylog} into the sum in eq.~\eqref{ladders}, with $2\sigma = \log{(z \bar{z})}$, and relabelling $j\rightarrow 2p-j$, one finds
\beq\label{eq:ladder-intermediate}
\begin{aligned}
L_p(z,\bar{z}) &= \sum_{j=0}^{p} \frac{(2p-j)!(-2\sigma)^{j}}{p!j!(p-j)!} [\text{Li}_{2p-j}(z) - \text{Li}_{2p-j}(\bar{z})] \\
&= \frac{2}{p! (p-1)!} \int\limits_{0}^{\infty} \rmd r \, r^{p-1}(r-\sigmam)(r-2\sigmam)^{p-1} \bigg[\frac{z}{e^{r}-z}-\frac{\bar{z}}{e^{r}-\bar{z}}\bigg] \\
&= \frac{(z-\bar{z})}{2\sqrt{z\bar{z}} \, p! (p-1)!} \int\limits_{-\sigmam}^{\infty} \frac{\rmd x x(x^2-\sigmam^2)^{p-1}}{\cosh{\tfrac{1}{2}(x+\sigmap)}\cosh{\tfrac{1}{2}(x-\sigmap)}}\, ,
\end{aligned}
\eeq
making use of the combinatorial identity (for $p>0$)
\beq
\sum_{j=0}^{p} \frac{(p-1)!(2p-j)!}{j!(p-j)!(2p-j-1)!}r^{p-j}(-2\sigmam)^{j} = 2(r-\sigmam)(r-2\sigmam)^{p-1} \, ,
\eeq
and changing variables, $x = r-\sigmam$. One recovers eq.~(\ref{int-ladd}) by manipulating the contour of integration, noting that the integrand is antisymmetric under $x\rightarrow -x$.

One then inserts eq.~\eqref{int-ladd} inside the determinant in eq.~\eqref{eq:I-det} and applies the so-called Cauchy-Binet-Andr\'eief formula (Lemma~\ref{lemma:andreief} in appendix \ref{sec:lemmas}) to bring the determinant below the integral sign.  Using eq.~\eqref{eq:IdualIrelation} to remove the factors of $(z-\bar{z})/\sqrt{z\bar{z}}$, one finds
\bea\label{eq:I-det-dual}
I^{{\rm Dual}}_{m, n} &= \frac{1}{2^m \C} \det_{1\leqslant i, j \leqslant m}\left[  \int_{|\sigmam|}^{\infty} \frac{\rmd x x(x^2-\sigmam^2)^{n-m+i+j-2}}{\cosh{\tfrac{1}{2}(x+\sigmap)}\cosh{\tfrac{1}{2}(x-\sigmap)}}\right] \\
&= \frac{1}{2^m m! \C}\prod_{\ell =1}^{m}\int_{|\sigmam|}^{\infty} \frac{\rmd x_{\ell} x_{\ell}(x_{\ell}^2-\sigmam^2)^{n-m}}{\cosh{\tfrac{1}{2}(x_{\ell}+\sigmap)}\cosh{\tfrac{1}{2}(x_{\ell}-\sigmap)}} [\Delta_{m}(x^2)]^2\,.
\eea
One encounters two copies of the determinant
\be
\Delta_{m}(x^2) \equiv \det_{1\leqslant i, j \leqslant m} \Bigl[ (x_{j}^2-\sigma^2)^{i-1} \Bigr] \,,
\ee
which is actually independent of $\sigma$, and is the Vandermonde determinant
\beq
\Delta_{m}(x^2) = \det_{1\leqslant i, j \leqslant m} (x_{j}^{2(i-1)})
= \prod_{j>i}^{m} (x^{2}_{j}-x^{2}_{i})\, ,
\eeq
associated with the set $\{x_{j}^2, j=1, \ldots , m\}$.\footnote{%
For an arbitrary set $\{y_{1}, \ldots , y_{N}\}$, the Vandermonde determinant is defined by $\Delta_{N}(y) = \det_{1\leqslant i, j \leqslant N} (y_{j}^{i-1}) = \prod_{j>i}^{N} (y_{j}-y_{i})$.}
Equation~\eqref{eq:I-det-dual} proves the equivalence of the dual and original representations, eqs.~\eqref{eqn-detInt} and~\eqref{eq:Idual}.

\subsection{BMN integral}\label{sec:BMN_det}

The first integrability-based integral to be discussed is the BMN integral in ref.~\cite{Basso:2017jwq}. This representation yields the pure function $I_{m, n}$ in the form of a sum of integrals,
\be
\begin{aligned}\label{eq:I_BMN}
I^{\rm BMN}_{m, n} &= \frac{1}{m!} \prod_{\ell=1}^{m}\SumInt_{\ell}   \frac{a_{\ell}\chi_{\ell}(z)}{(u_\ell^2 + a_\ell^2/4)^{m+n}} \prod_{i<j}^m  \left[(u_{i}-u_{j})^2 + \tfrac{1}{4}(a_{i}-a_{j})^2\right]\left[(u_{i}-u_{j})^2 + \tfrac{1}{4}(a_{i}+a_{j})^2\right]\, ,
\end{aligned}
\ee
with $\SumInt_{\ell} \equiv \sum_{a_{\ell}\geqslant 1} \int_{\mathbb{R}} \rmd u_{\ell}/(2\pi)$ and (with $\I = \sqrt{-1}$)
\beq
\chi_{\ell}(z) \equiv (z\bar{z})^{-\I u_{\ell}}\left((z/\bar{z})^{\frac{1}{2}a_{\ell}}-(\bar{z}/z)^{\frac{1}{2}a_{\ell}}\right) = 2(-1)^{a_{\ell}} \sinh{(a_{\ell} \sigmap)} e^{-2\I u_{\ell} \sigmam}\, .
\eeq
The translation to the variables $\sigma$ and $\varphi$ follows from eq.~\eqref{eq:z-mapping}, after fixing the square-root ambiguity appropriately. The representation~\eqref{eq:I_BMN} is initially defined in Euclidean kinematics, that is, for $\sigma \in \mathbb{R}$ and $-\I \varphi \in (-\pi, \pi)$, and then analytically continued. Its direct evaluation is straightforward for low values of $m$ and the result is seen to reproduce the ladder determinant. In particular, one easily verifies the agreement with the ladder series~\cite{Fleury:2016ykk},
\beq
L_{n}(z, \bar{z}) = \sum_{a_1=1}^{\infty} \int \frac{du_1}{2\pi} \frac{a_1 \chi_1(z)}{(u^2_1+\frac{a^2_1}{4})^{n+1}}\, ,
\eeq
when $m=1$. In this subsection, we shall prove the general identity
\beq\label{eq:goal-BMN}
I^{\rm BMN}_{m, n} = \left[\frac{z-\bar{z}}{\sqrt{z\bar{z}}}\right]^{m} I^{\rm Dual}_{m, n} = (e^{-\sigmap}-e^{\sigmap})^m I^{\rm Dual}_{m, n}\, ,
\eeq
with $I^{\rm Dual}_{m, n}$ the integral form of the determinant~\eqref{eq:Idual}, for all $m$ and $n$, with $n\geqslant m$.

To begin, we note that the bulk of the integrand in eq.~\eqref{eq:I_BMN} can be written concisely as a Vandermonde determinant. Namely, defining conjugate variables
\beq\label{eq:xi-var}
\xi_{2\ell-1}=u_\ell + \I a_\ell/2 \, , \qquad \xi_{2\ell}=u_\ell - \I a_\ell/2\, ,  \qquad \ell = 1, \ldots , m\, ,
\eeq
one immediately finds that the integrand can be cast into the compact form
\bea \label{eqn:sumBMN}
I^{\rm BMN}_{m, n} = \frac{\I^m}{m!}\prod_{\ell=1}^m \SumInt_{\ell}  \frac{\chi_{\ell}(z)}{(\xi_{2\ell-1}\xi_{2\ell})^{m+n}} \times \Delta_{2m}(\xi) \, ,
\eea
where $\Delta_{2m}(\xi)$ is the Vandermonde determinant for the set $\{\xi_{1}, \xi_{2}, \ldots , \xi_{2m}\}$,
\bea\label{eq:V2m}
\Delta_{2m}(\xi) &= \det\, (\xi_{j}^{i-1}) = \prod_{j>i}^{2m}(\xi_{j}-\xi_{i}) \\
&= \prod_{i=1}^{m} (-\I a_{i}) \times \prod_{i>j}^{m} \left[(u_{i}-u_{j})^2 + \tfrac{1}{4}(a_{i}-a_{j})^2\right]\left[(u_{i}-u_{j})^2 + \tfrac{1}{4}(a_{i}+a_{j})^2\right]\, .
\eea
One can make immediate use of this remarkable simplification to disentangle pairs of $\xi$ variables and cast the integral in the form of the Pfaffian of a $2m\times 2m$ skew-symmetric matrix $B$,
\beq
I^{\rm BMN}_{m, n} = \I^{m} \textrm{pf}\, B = \frac{\I^{m}}{2^{m} m!} \sum_{\pi \in S_{2m}} \sign{(\pi)} \prod_{\ell=1}^{m}B_{\pi(2\ell-1), \pi(2\ell)}\, ,
\eeq
with the sum running over the permutations of the set $\{1, \ldots , 2m\}$, by plugging
\beq\label{eq:delta-sum}
\Delta_{2m}(\xi) =\sum_{\pi \in S_{2m}} \sign{(\pi)} \prod_{\ell=1}^{m}(\xi_{2\ell-1}^{\pi(2\ell-1)-1}\xi_{2\ell}^{\pi(2\ell)-1})
\eeq
into eq.~\eqref{eq:V2m}, and defining
\beq\label{eq:B-int}
B_{ij} = \SumInt_{\ell} \frac{\chi_{\ell}(z)}{(\xi_{2\ell-1}\xi_{2\ell})^{m+n}} (\xi_{2\ell-1}^{i-1} \xi_{2\ell}^{j-1}-\xi_{2\ell-1}^{j-1} \xi_{2\ell}^{i-1})\, .
\eeq
A drawback is that the elements $B_{ij}(z, \bar{z})$ are not ladders, for $j\neq i\pm 1$, but derivatives thereof.%
\footnote{One finds, for $r\geqslant 1$,
\beq
B_{i,i+r}(z, \bar{z}) = -\frac{(\I z \partial_{z})^{r}-(\I \bar{z} \partial_{\bar{z}})^{r}}{z \partial_{z}-\bar{z} \partial_{\bar{z}}} L_{m+n-i}(z, \bar{z})\, .
\eeq
}
Further non-trivial algebraic identities are needed to ``purify'' the expression and reduce the integral to a determinant of an $m\times m$ matrix, as discussed in refs.~\cite{Kostov:2019stn,Kostov:2019auq,Belitsky:2019fan,Belitsky:2020qrm,Belitsky:2020qir,Kostov:2021omc} in the related context of large-charge correlators in $\mathcal{N}=4$ SYM.

Interestingly, it turns out to be possible to bypass this difficulty by decoupling all $\xi$ variables from the onset. (This promotes the full permutation symmetry of the integrand to the integral level, if not for the contours, see below.) The preliminary step is to relax the constraints $\I(\xi_{2\ell}-\xi_{2\ell-1}) = a_{\ell}\in \mathbb{N}$ among conjugate variables~\eqref{eq:xi-var}, by turning the sums over positive $a$'s into integrals. This task may be done efficiently with the help of the Mellin-Barnes summation technique. The key formula is Lemma~\ref{lemma:MBclassique} in appendix \ref{sec:lemmas}. It reads%
\footnote{A similar transformation was considered in ref.~\cite{Fleury:2016ykk} in relation to the Mellin representation of correlators.}
\bea\label{eq:MB}
\sum_{a =1}^{\infty} (-1)^{a } (e^{a \sigmap}-e^{-a \sigmap}) f(a) = \int\limits_{-\infty}^{\infty} \rmd \r \left[\frac{1}{e^{\sigmap-\r}+1} - \frac{1}{e^{-\sigmap -\r}+1}\right] \, \int_{\mathcal{C}} \frac{\rmd a}{2\pi \I} f(a) e^{-a \r}\, ,
\eea
for any suitably smooth function $f$ and with $\mathcal{C}$ parallel to the imaginary axis, $\mathcal{C} = \varepsilon + \I \mathbb{R}$ with $\varepsilon \in \left(0,1\right)$. This representation stems from the combination of the Sommerfeld-Watson transform,
\bea
\sum_{a=1}^{\infty} (-1)^{a} (e^{a \sigmap}-e^{-a \sigmap}) f(a) &= \I \int_{\varepsilon-\I\infty}^{\varepsilon+\I \infty} \rmd a \frac{\sinh{(a \sigmap)}}{\sin{(a \pi)}} f(a)\, ,
\eea
with an integral transform for the inverse sine factor, see eq.~\eqref{eq:int-sine} in appendix~\ref{sec:lemmas}. This combination is well-suited to our problem as it puts in place the dual variable $x$ entering the integral form of the ladder determinant. It also removes the constraints on $\sigmap$, which may now be chosen anywhere inside the extended kinematics $\textrm{Im}\, \sigmap \in (-\pi, \pi)$. Applying eq.~\eqref{eq:MB} to each sum in eq.~\eqref{eqn:sumBMN}, we get
\bea
I^{\rm BMN}_{m, n} & = \frac{(-\I)^m}{2^m m!}   \int_{\mathbb{R}^m}\prod_{\ell=1}^m \frac{\rmd \r_{\ell}(e^{\sigmap}-e^{-\sigmap})}{2\cosh{\frac{1}{2}(\r_{\ell}+\sigmap)}\cosh{\frac{1}{2}(\r_{\ell}-\sigmap)}}\,  \mathcal{I}_{1}(\sigmam, \{x\}) \, ,
\eea
with all dependence on $\sigmam$ factorizing neatly into the integral
\beq
\mathcal{I}_{1}(\sigmam, \{x\}) = \prod_{\ell=1}^{m}\int_{\mathcal{C}} \frac{\rmd a_\ell}{2\pi \I} \int_{\mathbb{R}} \frac{\rmd u_{\ell}}{2\pi} \frac{e^{-a_{\ell} x_{\ell} -2\I u_{\ell}\sigmam}}{(\xi_{2\ell-1}\xi_{2\ell})^{m+n}} \times \Delta_{2m}(\xi)\, .
\eeq
The key simplification is that $\xi_{2\ell-1}$ and $\xi_{2\ell}$ can now be regarded as independent variables. They would be real for $u_{\ell} \in \mathbb{R}$ and $a_{\ell}\in \I \mathbb{R}$. Instead, they run here parallel to the real axis, along the contours
\beq
\xi_{2\ell-1} \in \Gamma_{+} \equiv \mathbb{R} + \I \varepsilon\, , \qquad \xi_{2\ell} \in \Gamma_{-}\equiv \mathbb{R} - \I \varepsilon\, ,
\eeq
because $a_{\ell}$ has a small positive real part $\varepsilon$. Hence,
\beq
\int_{\mathcal{C}} \frac{\rmd a_\ell}{2\pi \I} \int_{\mathbb{R}} \frac{\rmd u_{\ell}}{2\pi} =  \int_{\Gamma_+ \times \Gamma_-} \frac{\rmd  \xi_{2\ell-1} \rmd  \xi_{2\ell}}{(2\pi \I )^2} \, , \qquad \forall \ell= 1, \ldots , m\, ,
\eeq
after taking into account the Jacobian ($ = -\I$) for the change of variables~(\ref{eq:xi-var}). We may now write $\mathcal{I}_{1}(\sigma, \{x\})$ as a bunch of decoupled $\xi$-integrals using eq.~(\ref{eq:delta-sum}) and recast it as the determinant of a $2m\times 2m$ matrix $A$,
\bea
\label{eq:detColumns}
\mathcal{I}_{1}(\sigmam, \{x\})  = \det\,  A\, ,
\eea
with elements,
\beq
A_{k, 2\ell-1} = A^{+}_{k}(\r_{\ell})\, , \qquad A_{k, 2\ell} = A^{-}_{k}(\r_{\ell})\, , \qquad k =1, \ldots , 2m,  \,\,\, \ell = 1, \ldots , m\, ,
\eeq
and
\beq\label{eq:ind-int}
A^{\pm}_k(\r) \equiv \int_{\Gamma_\pm} \frac{\rmd \xi}{2\pi \I} \xi^{k-m-n-1} e^{\I (\pm \r-\sigmam) \xi}\, .
\eeq
Closing the contour of integration in the upper/lower half $\xi$ plane, depending on whether $\pm x-\sigmam \gtrless 0$, one finds
\bea\label{eq:int-phi}
A^{\pm}_k(\r)
=& \mp\frac{1 }{(m+n-k)!} [\I(\pm \r - \sigmam) ]^{m+n-k} \theta(-\r \pm \sigmam)\, ,
\eea
with $\theta(t)$ the Heaviside step function, $\theta(t) =1$ for $t\geq 0$ and $\theta(t) = 0$ otherwise.

One proceeds with straightforward linear algebra. Plugging~\eqref{eq:int-phi} into eq.~\eqref{eq:detColumns}, we first factor out $\mp [\I(\pm \r_\ell - \sigmam) ]^{n-m} \theta(-\r_\ell \pm \sigmam)$, column by column, and then $1/(m+n-k)!$, row by row. The leftover factor is a Vandermonde determinant for the set $\{ \I(\r_1-\sigma), \I(- \r_1-\sigma), \I(\r_2-\sigma), \I(-\r_2-\sigma),\dots, \I(\r_m-\sigma), \I(-\r_m-\sigma)\}$. As such, it does not depend on $\sigmam$ and can be written concisely as
\beq
(-\I)^{m} \prod_{i<j}^{m} (\r_i-\r_j)^2 \prod_{i,j}^{m} (\r_i + \r_j) = (-\I)^{m} \prod_{i=1}^{m} 2\r_{i} \times [\Delta_m(\r^2)]^2\, .
\eeq
Assembling all factors together, we obtain
\bea\label{eq:simp}
\mathcal{I}_{1}(\sigmam)
&= \frac{\I^{m}}{\mathcal{N}} \prod_{i=1}^m  \theta(  -|\sigmam|-\r_i) 2\r_i (\r_i^2 - \sigmam^2)^{n-m} \times [\Delta_m(\r^2)]^2\, ,
\eea
and, therefore, 
\beq
I^{\rm BMN}_{m, n} = \frac{(e^{\sigmap}-e^{-\sigmap})^{m}}{2^{m} m! \mathcal{N}}  \prod_{i=1}^m\int_{-\infty}^{-|\sigma|} \frac{ \rmd x_{i} \r_i (\r_i^2 - \sigmam^2)^{n-m}}{\cosh{\frac{1}{2}(\sigmap+x_{i})}\cosh{\frac{1}{2}(\sigmap-x_{i})}} [\Delta_m(\r^2)]^2\, ,
\eeq
which implies eq.~\eqref{eq:goal-BMN} with eq.~\eqref{eq:I-det-dual} and concludes the proof that the BMN integral~\eqref{eq:I_BMN} is equivalent to the determinant formula~\eqref{eq:I-det}.

It is worth mentioning that the above analysis would also apply to the anisotropic fishnet diagrams \cite{Zamolodchikov:1980mb,Kazakov:2018qbr} associated with a theory in which the two scalar fields are given non-canonical dimensions $1\mp 2\omega$, with $\omega \in [0,1/2)$ playing the role of an anisotropy parameter. The equivalent of the ``BMN representation'' for the $m\times n$ deformed fishnet diagram in four dimensions was worked out in refs.~\cite{Derkachov:2019tzo,Derkachov:2020zvv}, along with other generalizations of the correlator, using the method of separation of variables. It was found to take the same form as for isotropic fishnets, that is~\eqref{eq:I_BMN}, up to the replacement
\beq
\frac{1}{(u_{\ell}^2 + a_{\ell}^2/4)^{m+n}} \rightarrow [f(\tfrac{a_{\ell}}{2}-\I u_{\ell}) f(\tfrac{a_{\ell}}{2}+\I u_{\ell})]^{m+n}, 
\eeq 
with $f(y) = \Gamma(y+\omega)/\Gamma(y+1-\omega)$ and with $\Gamma$ the Euler Gamma function.%
\footnote{Incidentally, similar generalizations appear in the context of the exact solutions to the Kardar-Parisi-Zhang equation in 1+1 dimensions, where the initial geometry provides the anisotropy, see e.g.~refs.~\cite{imamura2012exact, borodin2015height, barraquand2020half} for instances where $f$ is a ratio of Gamma functions.} Importantly, since the $\xi$-integrals stay factorized under this deformation, one can repeat the above analysis and derive a $2m\times 2m$ determinant representation for the correlator by slightly modifying the integrals in eq.~\eqref{eq:ind-int}, using
\begin{equation}
A_{k}^{\pm}(\r) \rightarrow \int_{\Gamma_{\pm}}\frac{d\xi}{2\pi \I}\xi^{k-1}[\mp \I f(\mp \I \xi)]^{m+n}  e^{\I (\pm \r-\sigmam) \xi}\, .
\end{equation}
It is not clear, however, if the result may be further simplified and cast into the form of a determinant of an $m\times m$ matrix, as in the undeformed case, since the factorization of the determinant observed in eq.~\eqref{eq:simp} depends very much on the explicit form of the $\xi$-integrals. In contrast, 2d fishnet correlators were found \cite{Derkachov:2018rot} to admit a uniform $m\times m$ determinant representation for \textit{all} values of the anisotropy parameter.

\subsection{FT integral}\label{sec:FT_det}

\begin{figure}[t]
\begin{minipage}[c]{0.5\linewidth}
\centering
\includegraphics[width=0.6\textwidth]{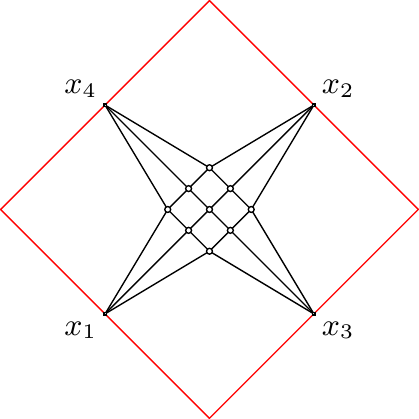}
\end{minipage}
\hspace*{-0.5cm}
\begin{minipage}[c]{0.5\linewidth}
\centering
\includegraphics[width=0.6\textwidth]{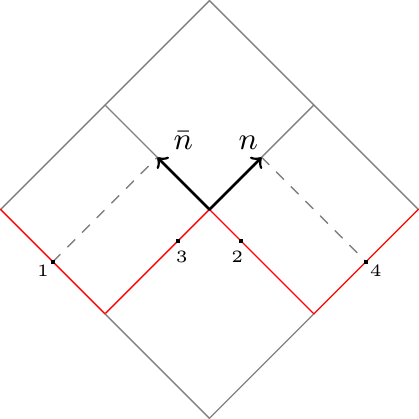}
\end{minipage}
\caption{The flux-tube factorization relies on lightcone coordinates to parametrize the positions of the four points $x_{1,2,3,4}$ along a null conformal frame (square). In the case of interest, i.e.~for spacelike separations, $x_{ij}^2 <0$, the null square is the Poincar\'e patch of (conformally compactified) $\mathbb{R}^{1,1} \subset \mathbb{R}^{1,3}$, as shown in the right panel. Two points are placed at the boundary of the patch, with $n\cdot x_{1} = \bar{n}\cdot x_{4} = 1$ and $n, \bar{n}$ two orthogonal lightlike vectors, $n^2 = \bar{n}^2 = 0$, $n\cdot \bar{n} = 1$. The remaining points are sitting at $n\cdot x_{2} = z$ and $\bar{n}\cdot x_{3} = \bar{z}$, with $z, \bar{z}$ both real and negative.}\label{FT-figure}
\end{figure}

The second integrability-based representation is the so-called flux-tube representation. It arises in the Minkowskian kinematics when one considers the correlator as sitting inside a null conformal frame, as depicted in figure \ref{FT-figure}. The end-points of the correlator are then interpreted as producing beams of flux-tube particles at the positions
\beq\label{eq:map-sigma}
z = -e^{2\sigma_{1}} = -e^{\sigmap+\sigmam}\, , \qquad \bar{z} = -e^{-2\sigma_{2}} = -e^{\sigmam-\sigmap}\, ,
\eeq
with $\sigma_{1,2} \in \mathbb{R}$, along two orthogonal lightcone directions, $n^2= \bar{n}^2 =0$, $n\cdot \bar{n}=1$, as explained in the caption of figure \ref{FT-figure}.

A nice feature of the flux-tube representation is that it treats symmetrically the $m$ and $n$ lines of the fishnet graph. Each set of lines is assigned to a set of rapidities, $\{u_{i=1, \ldots, m}\}$ and $\{v_{j=1, \ldots , n}\}$, representing the momenta of the flux-tube excitations, moving across the conformal square, and conjugate to the coordinates \eqref{eq:map-sigma}. The price to pay is that it gives rise to more complicated integrals involving both rational and hyperbolic functions. As explained in ref.~\cite{Basso:2017jwq}, in this representation, the correlator takes the form
\be \label{eqn:ftRelation}
\begin{aligned}
\Phi_{m,n}
& = d(z, \bar{z})^m \times \pi^{\frac{1}{2}(m+n)(m+n+1)}(2\cosh \sigma_2)^{n-m}\times  I^{\rm FT}_{m, n}\, ,
\end{aligned}
\ee
where $I^{\rm FT}_{m, n}$ is the $m\times n$ Fourier integral%
\footnote{We simplified the original expression in ref.~\cite{Basso:2017jwq} by using $\sinh{(\pi(x-y))} = \cosh{(\pi x)} \cosh{(\pi y)}\, (\tanh{(\pi x)} - \tanh{(\pi y)})$ and stripping off powers of $\pi$.}
\be\label{eqn-FT}
\begin{split}
&I^{\rm FT}_{m, n}=
\int\limits_{\mathbb{R}^{m}}\frac{\rmd \bm{u}}{m!} \int\limits_{\mathbb{R}^{n}}\frac{\rmd \bm{v}}{n!} \prod_{i=1}^m\frac{e^{2\I u_i \sigma_1}}{\cosh{(\pi u_i)}}\prod_{j=1}^n\frac{e^{2\I v_j \sigma_2}}{\cosh{(\pi v_j)}}  \\
&\qquad \qquad \times \Delta_m(u)\Delta_m(\tanh{(\pi u)})\Delta_n(v)\Delta_n(\tanh{(\pi v)})\prod_{i,j}^{m, n}\frac{\tanh{(\pi u_i)} - \tanh{(\pi v_j)}}{u_i - v_j}\, ,
\end{split}
\ee
with $\rmd \bm{u} = \prod_{i=1}^{m}du_{i}/(2\pi), \rmd \bm{v}= \prod_{j=1}^{n}dv_{j}/(2\pi),$ and with the $\Delta$'s referring as before to Vandermonde determinants,
\beq
\Delta_{m}(u) = \prod_{j>i}^{m}(u_{j}-u_{i})\, , \qquad \Delta_{m}(\tanh{(\pi u)}) = \prod_{j>i}^{m}(\tanh{(\pi u_{j}})-\tanh{(\pi u_{i})})\, ,
\eeq
and similarly for the $v$'s. Integrals of this type were discussed in refs.~\cite{Derkachov:2016dhc,Derkachov:2019ynh,Derkachov:2020zvv} in connection with the separation of variables for $SL(2, \mathbb{R})$ spin chains. In particular, in the cases where $m=0$ or $n=0$, the integral matches with the norm of the separated wave function, and it can be evaluated straightforwardly using eq.~(54) of ref.~\cite{Derkachov:2016dhc}. Below we explain how to calculate the integral in the general case, $m, n \neq 0$.

The mixing among hyperbolic and rational interactions makes the flux-tube integral harder to treat than the BMN one. The problem can be circumvented, however,  by collecting these two types of interactions into suitable determinants.

The rational part, for instance, can be encoded in a generalized Cauchy determinant \cite{Han:2000rec},
\beq
\frac{\Delta_m(u) \Delta_n(v)}{\prod_{i,j} (u_i - v_j)} =  (-1)^{\frac{1}{2}m(2n-m-1)}  \times \det_n C\, ,
\eeq
where $C$ is a ``centaur'' $n\times n$ matrix, mixing Cauchy and Vandermonde entries, defined for $n\geqslant m$ and two sets of variables $\{u_{1}, \ldots , u_{m}\}$ and $\{v_{1}, \ldots , v_{n}\}$ by
\be
C=
\left(
\phantom{
\begin{matrix}
 \dfrac{1}{u_1 - v_n}  \vspace{0.5em}\\  \dfrac{1}{u_2 - v_n} \\ \vdots\\  \dfrac{1}{u_m - v_n}\\  \\  \\ 1\\ v_n\\  \vdots\\  v_n^{n-m-1} 
\end{matrix}}
\right.
\hspace{-3.5em}
\begin{matrix}
\dfrac{1}{u_1 - v_1} & \dfrac{1}{u_1 - v_2} & \cdots &  \dfrac{1}{u_1 - v_n} \vspace{0.5em} \\
\dfrac{1}{u_2 - v_1} & \dfrac{1}{u_2 - v_2} & \cdots &  \dfrac{1}{u_2 - v_n} \\
\vdots & \vdots & \vdots & \vdots \\
\dfrac{1}{u_m - v_1} & \dfrac{1}{u_m - v_2} & \cdots &  \dfrac{1}{u_m - v_n} \\
&&&\vspace{-0.5em} \\
 \hline \vspace{-1em} \\
1 & 1 & \cdots & 1 \\
v_1 & v_2 & \cdots & v_n \\
\vdots & \vdots & \vdots & \vdots \\
v_1^{n-m-1} & v_2^{n-m-1} & \cdots & v_n^{n-m-1} 
\end{matrix}
\hspace{-3.5em}
\left.\phantom{
\begin{matrix}
 \dfrac{1}{u_1 - v_n}  \vspace{0.5em}\\  \dfrac{1}{u_2 - v_n} \\ \vdots\\  \dfrac{1}{u_m - v_n}\\  \\  \\ 1\\ v_n\\  \vdots\\  v_n^{n-m-1} 
\end{matrix}}
\right)
\hspace{-0em}
\begin{tabular}{l}
$\left.\lefteqn{\phantom{\begin{matrix} 
   \dfrac{1}{u_1 - v_n}  \vspace{0.5em}\\ \dfrac{1}{u_2 - v_n} \\ \vdots\\  \dfrac{1}{u_m - v_n}\\   \hline \vspace{-1em} \\ ~\ 
\end{matrix}}}\right\}m$\\
$\left.\lefteqn{\phantom{\begin{matrix}   1\\ v_n\\  \vdots\\ v_n^{n-m-1} \\ ~\
\end{matrix}}} \right\}n-m$
\end{tabular} \, .
\ee
The hyperbolic factors, on the other hand, can be seen as coming from the action of the Vandermonde differential operator
\beq
\Delta_{m+n}(\{\overrightarrow{\p_{v_j}},\overrightarrow{\p_{u_i}}\}) \equiv \prod_{i>j}^{m} (\p_{u_{j}}-\p_{u_{i}}) \prod_{i>j}^{n} (\p_{v_{j}}-\p_{v_{i}}) \prod_{i, j}^{n, m} (\p_{v_{j}}-\p_{u_{i}})
\eeq
on the factor $\prod_{i,j} \sech({\pi u_i)} \sech{(\pi v_j)}$, as shown in appendix \ref{sec:lemmas} (Lemma~\ref{lemma:vandermonde}). Accordingly, the integral can be written more concisely as
\bea\label{eq:FT-dets}
I^{\rm FT}_{m, n}  & =\frac{(-1)^{\frac{1}{2}m(2n-m-1)}}{\prod_{j=1}^{m+n} (j-1)!}\left(\frac{-1}{\pi}\right)^{\frac{1}{2}(m+n)(m+n-1)} \\
& \times \int_{\mathbb{R}^{m}}\frac{\rmd \bm{u}}{m!} \int_{\mathbb{R}^{n}}\frac{\rmd \bm{v}}{n!}\prod_{i=1}^me^{2\I u_i \sigma_1}\prod_{j=1}^n e^{2\I v_j \sigma_2} \times \det\limits_n C \times \Delta_{m+n}(\{\overrightarrow{\p_{v_j}},\overrightarrow{\p_{u_i}}\}) \prod_{i,j} \sech({\pi u_i)} \sech{(\pi v_j)} \,.\\
\eea
We may then perform an integration by parts, using
\beq
\Delta_{m+n}(\{\overrightarrow{\p_{v_j}},\overrightarrow{\p_{u_i}}\}) = \Delta_{m+n}(\{-\overleftarrow{\p_{v_j}},-\overleftarrow{\p_{u_i}}\}) = (-1)^{\frac{1}{2}(m+n)(m+n-1)}\Delta_{m+n}(\{\overleftarrow{\p_{v_j}},\overleftarrow{\p_{u_i}}\})\, , 
\eeq
with the arrows indicating on which side of the integrand in eq.~(\ref{eq:FT-dets}) the derivative is acting. Note that there are no boundary terms, since the integrand is exponentially small at infinity. Using the Leibniz formula for multi-linear forms,
\bea
\label{eq:Leibniz}
&\Delta_{m+n}(\{\overrightarrow{\p_{v_j}},\overrightarrow{\p_{u_i}}\}) \bigg [e^{2\I \sum_{i}u_i \sigma_1+2\I \sum_{j}v_j \sigma_2}  \mathcal{F}(\{u_{i},v_{j}\})\bigg] \\
&\qquad \qquad =e^{2\I \sum_{i}u_i \sigma_1+2\I \sum_{j}v_j \sigma_2}  \Delta_{m+n}(\{\overrightarrow{\p_{v_j}}+2\I \sigma_2,\overrightarrow{\p_{u_i}}+2\I \sigma_1\})\mathcal{F}(\{u_{i},v_{j}\})\, ,
\eea
which applies to any function $\mathcal{F}(\{u_{i},v_{j}\})$, we get
\bea\label{eq:intermediate}
I^{\rm FT}_{m, n}  & = \frac{(-1)^{\frac{1}{2}m(2n-m-1)}}{\prod_{j=1}^{m+n} (j-1)!} \left(\frac{1}{\pi}\right)^{\frac{1}{2}(m+n)(m+n-1)} \\
& \times \int_{\mathbb{R}^{m}}\frac{\rmd \bm{u}}{m!} \int_{\mathbb{R}^{n}}\frac{\rmd \bm{v}}{n!}\prod_{i=1}^m\frac{e^{2\I u_i \sigma_1}}{\cosh{(\pi u_i)}}\prod_{j=1}^n \frac{e^{2\I v_j \sigma_2}}{\cosh{(\pi v_j)}}  \times \Delta_{m+n}(\{\overrightarrow{\p_{v_j}}+2\I \sigma_2,\overrightarrow{\p_{u_i}}+2\I \sigma_1\})\det_n C\, .
\eea

The next step is to disentangle the integrals in $\{u_{i}\}$ and $\{v_{j}\}$. We may do so by exponentiating the first $m$ lines of the matrix $C$ using Schwinger's trick,
\beq
\frac{1}{u-v} = -\I \int_0^\infty \rmd r \, e^{\I r(u-v + \I 0)}\, ,
\eeq
with $\I0$'s introduced to ensure convergence. We shall omit the latter prescription in the following, since the integrand is regular when $u_i=v_j, \forall i, j$. Factoring out $e^{\I r_i u_i}$, row by row, we get
\begin{equation}\label{eq:CtildeC}
\det_n C= (-\I)^m \int_{\mathbb{R}^m_+}\rmd \vec{r} \, e^{\I \sum_{i}r_i u_i} \det_n \tilde{C}\, ,
\end{equation}
with $\rmd \vec{r} = \prod_{i=1}^{m} \rmd r_{i}$, $\mathbb{R}_{+} = (0, \infty)$,  and where the matrix $\tilde{C}$ has the same structure as $C$ up to the replacement $1/(u_i-v_j)\to e^{-\I r_i v_j}$. Plugging~\eqref{eq:CtildeC} into \eqref{eq:intermediate} and applying the Leibniz formula to the phases in eq.~\eqref{eq:CtildeC} allow us to strip off the $\lbrace u_i \rbrace $ dependence,
\bea
I^{\rm FT}_{m, n}  & = \frac{(-\I)^{m(2n-m)}}{\prod_{j=1}^{m+n} (j-1)!}\left(\frac{1}{\pi}\right)^{\frac{1}{2}(m+n)(m+n-1)}\\
& \times \int_{\mathbb{R}^m_+}\frac{\rmd \vec{r}}{m!} \, \underbrace{ \int_{\mathbb{R}^{m}}  \rmd \bm{u}  \prod_{i=1}^m\frac{e^{\I(2 \sigma_1 + r_i )u_i }}{\cosh{(\pi u_i)}}}_{\displaystyle \mathcal{I}_2}  \int_{\mathbb{R}^{n}} \frac{\rmd \bm{v}}{n! } \prod_{j=1}^n \frac{e^{2\I v_j \sigma_2}}{\cosh{(\pi v_j)}}  \underbrace{\Delta_{m+n}(\{\overrightarrow{\p_{v_j}}+2\I \sigma_2,\I r_i+2\I \sigma_1\}) \det_n \tilde{C}}_{\displaystyle \mathcal{I}_3}\, .
\eea
The integral $\mathcal{I}_{2}$ is fully factorized, so it can be evaluated directly using a Fourier transform,
\beq
\mathcal{I}_2=\prod_{i=1}^m\frac{1}{2\pi \cosh{(\sigma_1+r_i /2)}}\, .
\eeq
We then treat the action of the Vandermonde differential operator on $\textrm{det}\, \tilde{C}$. We find
\begin{equation}
\mathcal{I}_3= \I^{\frac{1}{2}m(m-1)} \Delta_m(r)\Delta_n(\partial_v) \prod_{j=1}^{n} P(\p_{v_j})\det_n \tilde{C}\, ,
\end{equation}
after using $\Delta_m(\I(r+2\sigma_1))=  \I^{\frac{1}{2}m(m-1)}\Delta_m(r), \Delta_n(\partial_v+2\I \sigma_2)=\Delta_n(\partial_v)$, and defining the polynomial
\beq
P(X) \equiv \prod_{j=1}^m (2\I \sigmam+\I r_j-X)\, ,
\eeq
with $\sigma = \sigma_{1}-\sigma_{2}$, see eq.~\eqref{eq:map-sigma}. Acting then on the columns of $\tilde{C}$ with $P(\p_{v_j})$ and collecting factors row by row, we find
\beq
\prod_{j=1}^n P(\partial_{v_j})\det_n \tilde{C}=P(0)^{n-m}\prod_{i=1}^m P(-\I r_i) \times \det_n \tilde{C}\, .
\eeq
We proceed with the computation of the $v$-integrals,
\bea
\label{eq:IFTintermediate}
I^{\rm FT}_{m, n} & = \frac{(-\I)^{\frac{1}{2}m(2n-m+1)}}{\prod_{j=1}^{m+n} (j-1)!} \left(\frac{1}{\pi}\right)^{\frac{1}{2}(m+n)(m+n-1)} \\
& \qquad  \times \int_{\mathbb{R}^m_+}\frac{\vec{\rmd r}}{m!} \, \prod_{i=1}^m\frac{(r_{i}+2\sigma)^{n-m}P(-\I r_{i})}{2\pi \cosh(\sigma_1+\frac{r_i}{2})} \Delta_m(r)   \underbrace{\int_{\mathbb{R}^{n}} \frac{\rmd \bm{v}}{n! } \prod_{j=1}^n \frac{e^{2\I v_j \sigma_2}}{\cosh{(\pi v_j)}}\Delta_n(\partial_v) \det_n \tilde{C}}_{\displaystyle \mathcal{I}_4}\, .
\eea
Applying the Cauchy-Binet-Andr\'eief formula  (Lemma~\ref{lemma:andreief}) with the measure $\rmd \nu = \rmd v \, e^{2\I v \sigma_2} \sech(\pi v)/(2\pi)$, and the function $g_j(v)=e^{-\I r_j v}$ for $j\in [1,m]$ and $g_{j+m}(v)=v^{j-1}$ for $j\in [1,n-m]$, we obtain
\beq
\mathcal{I}_4=\det_{n}\left[\int_\mathbb{R}\frac{\rmd v \, e^{2\I v \sigma_2}}{2\pi \cosh{(\pi v)}} \p_v^{k-1}  g_j(v)\right] = \det_{1\leqslant j, k \leqslant n} N_{jk}\, .
\eeq
The $n\times n$ matrix $N$ so defined has the special structure
\beq
N = \begin{pmatrix}
D & F \\ E & 0
\end{pmatrix}\, ,
\eeq
where the matrix $D$ is irrelevant to $\det N$, $E$ is a lower-triangular $(n-m)\times (n-m)$ matrix with diagonal elements
\beq
E_{jj} =(j-1)! \int_\mathbb{R}\frac{\rmd v\, e^{2\I \sigma_2 v}}{2\pi \cosh \pi v} = \frac{(j-1)!}{2\pi \cosh \sigma_2} \, ,
\eeq
and $F$ is an $m\times m$ Vandermonde matrix with elements
\beq
F_{jk}=\int_\mathbb{R}\frac{\rmd v\, e^{\I (2 \sigma_2- r_j)v}}{2\pi \cosh{(\pi v)}}(-\I r_j)^{k+n-m-1} =\frac{(-\I r_j)^{k+n-m-1}}{2\pi \cosh(\sigma_2-\frac{r_j}{2})}\, .
\eeq
Calculating $\mathcal{I}_{4}$ using the block determinant formula $\mathcal{I}_4= (-1)^{m(n-m)} \det E \det F$, along with the sub-determinants,
\begin{equation}\label{eq:IntermDeter}
\det_{n-m} E=\prod_{j=1}^{n-m} \frac{(j-1)!}{2\pi \cosh  \sigma_2}, \qquad \det_m F=(-\I)^{\frac{1}{2}m(2n-m-1)} \prod_{j=1}^m \frac{r_j^{n-m}}{2\pi \cosh(\sigma_2-\frac{r_j}{2})}\times \Delta_m(r)\, ,
\end{equation}
and collecting the numerous constant factors, we finally get
\begin{equation}
I^{\rm FT}_{m, n} = \frac{(-\I)^{m^2}}{(2\cosh{\sigma_{2}})^{n-m}}\left(\frac{1}{\pi}\right)^{\frac{1}{2}(m+n)(m+n+1)} \int_{\mathbb{R}^m_+} \frac{\rmd \vec{r}}{2^m m!\mathcal{N}} \prod_{i=1}^m \frac{r_i^{n-m}(r_{i} + 2\sigmam)^{n-m}P(-\I r_{i})}{2\cosh(\sigma_2-\frac{r_i}{2})\cosh (\sigma_1+\frac{r_i}{2})}[\Delta_m(r)]^2\, .
\end{equation}
We may now conclude. Plugging $P(-\I r_i)=\I^m \prod_{j=1}^m (2\sigmam+r_j+r_i)$ into the above expression and noting that
\begin{equation}
\prod_{i=1}^m P(-\I r_i) [\Delta_m(r)]^2=\I^{m^2} \big[\Delta_m\big((r+\sigmam)^2\big)\big]^2 \prod_{i=1}^m (2r_i +2 \sigmam)\, ,
\end{equation}
we obtain the sought-after relation
\beq
I^{\rm FT}_{m, n} = \frac{1}{(2\cosh{\sigma_{2}})^{n-m}}\left(\frac{1}{\pi}\right)^{\frac{1}{2}(m+n)(m+n+1)} I^{\textrm{Dual}}_{m, n}\, ,
\eeq
after changing variables, $r_{i} = x_{i}-\sigma$, and using eq.~\eqref{eq:map-sigma}.  We conclude that the flux-tube representation~\eqref{eqn-FT} is equivalent to the dual representation~\eqref{eq:I-det-dual}.

\section{Thermodynamic limit} \label{sec:thermodynamic}

We now turn to the thermodynamic limit of the rectangular fishnet correlator, which we define as the limit $n, m \rightarrow \infty$, holding fixed the aspect ratio of the rectangle, $k \equiv n/m \in (1, \infty)$. We will see that the correlator scales properly in this limit, with a finite ``free energy'' per site,%
\footnote{Note that we do not include a minus sign in the definition of the free energy, for convenience.}
\beq
F(k) \equiv \lim\limits_{m, n \rightarrow \infty} \frac{1}{mn}\log{\Phi_{m,n}}\,.
\eeq
The free energy is independent of the cross ratios ($z,\zb$ or $\varphi,\sigma$), as long as they are held fixed at generic values, but it is a nontrivial function of the aspect ratio $k$. This analysis can be done by applying the large $m$ saddle-point method to either of the integrals introduced earlier. However, an analytic form for the free energy $F$ is more easily found if one combines several solutions together. We analyze here two of the previously discussed integrals, the dual and BMN integrals, leaving aside the flux-tube integral which is substantially harder to study.

\subsection{Dual integral}

As seen earlier, the determinant formula~\eqref{eq:Phi-I} can be recast as a matrix-model-like integral~\eqref{eqn-detInt}, or
\beq\label{Jmn}
\Phi_{m, n} =  \frac{d(z, \bar{z})^m}{2^{m} m! \, \C} \int\limits_{|\sigmam|}^{\infty} \rmd x_{1} \ldots \rmd x_{m} \times e^{-\Gamma (\{x_{i}\})} \, ,
\eeq
with the prefactor defined in eq.~\eqref{Gmn} and with the effective potential
\beq\label{eq:eff-potential}
\Gamma \equiv m \sum_{i = 1}^{m} V(x_{i}) - \sum_{i\neq j}^{m} \log{|x^2_{i}-x^2_{j}|}\, .
\eeq
The latter comprises the Vandermonde interactions and the potential
\beq\label{eq:V}
V(x) \equiv -\frac{1}{\beta} \log(x^2-\sigmam^2) -\frac{1}{m}\log{x} +\frac{1}{m} \log\Bigl[\cosh{\tfrac{1}{2}(x+\sigmap)}\cosh{\tfrac{1}{2}(x-\sigmap)}\Bigr] \, ,
\eeq
where
\beq
\beta \equiv \frac{m}{n-m} = \frac{1}{k-1}\, ,
\eeq
and $k = n/m$. This integral is similar to known matrix models, such as the $O(-2)$ model, see~\cite{Kostov:1988fy,Gaudin:1989vx,Kostov:1992pn} and references therein, and is readily amenable to the large $m$ saddle-point method.

The key simplification at large $m$ is that the effective potential scales large, $\Gamma = \mathcal{O}(m^2)$, and thus the integral is sharply peaked at its minimum, $0 = \delta \Gamma / \delta x_{i}$. The latter equations can be interpreted as the conditions for the static equilibrium of a symmetric collection of charges at positions $\{\pm x_{i}\}$ subject to Coulomb-like interactions in the presence of the external potential $V(x)$.

In this regime, only the first and last terms on the right-hand side of eq.~\eqref{eq:V} remain. The latter is superficially small, but should nonetheless be kept, since it ensures the convergence of the integral at large $x$. Without it, the charges would run away, $x_{i} \rightarrow \infty$. Instead, their course stops at large distance $\sim m$, owing to the linear scaling at large $x$, $V(x) \approx |x|/m$, which in turn determines the large $m$ scaling,
\beq\label{eq:scaling-x}
x_{i} = \mathcal{O}(m)\, .
\eeq
The first term in eq.~\eqref{eq:V} is of order $\mathcal{O}(m^0)$, for $\beta \in (0, \infty)$, and it acts at the lower bound of the domain. It generates a logarithmic repulsion, which may be interpreted as coming from $n-m$ non-dynamical charges which push the dynamical charges away from $x^2 = \sigma^2$. Altogether, we expect that the charges are, after rescaling, densely distributed on a compact domain,
\beq
x_i/m \in \mathcal{C} \subset (|\sigma|/m, \infty)\, ,
\eeq
when $\beta$ is finite.%
\footnote{The situation is subtle when $n - m \rightarrow 0$, since the charges are then free to move all the way to $x = 0$ where they accumulate. We shall only consider this regime as the limit $k = n/m \rightarrow 1$ of the large $m, n$ solution.}
Furthermore, it readily follows from eq.~\eqref{eq:scaling-x} that the spacetime parameters $\sigmap, \sigmam$ drop out in the large $m$ limit, as long as $\sigmap, \sigmam$ stay of order $\mathcal{O}(1)$ at large $m$. Hence, one may approximate the potential by
\beq\label{eq:scaledV}
V \approx -\frac{1}{\beta} \log{x^2} +\frac{|x|}{m}\, ,
\eeq
in the thermodynamic limit discussed here. A more general scaling where the spacetime parameters scale large with $m$ will be discussed later on, in section~\ref{sec:spinning}.

\subsubsection{Saddle-point equation}

We may now develop the standard analysis and introduce a density
\beq\label{constraint}
\rho(x) = \frac{\beta}{2m}
\sum_{i=1}^{m} \delta\Big( x-\frac{\beta x_{i}}{4\pi m} \Big) \, ,
\eeq
which we normalized for convenience such that
\beq\label{normalisation}
\int_{\mathcal{C}}\rmd x \, \rho(x) = \frac{\beta}{2}\, .
\eeq 
We expect it to admit a smooth description at large $m$ over the support of the distribution $\mathcal{C}$. Since the scaled potential~\eqref{eq:scaledV} is strictly convex downward and diverges at the origin, the contour is a single interval $\mathcal{C} = (a, b)$ with $0<a\leqslant b <\infty$, for finite $\beta$. In particular, for a small interval $b\sim a$, the charges fill the bottom of the potential described by the Gaussian matrix model. It corresponds to the low density regime, $\beta \sim 0$, that is, $k \rightarrow \infty$. In the opposite regime, when $k \rightarrow 1$, the logarithmic repulsion at $x=0$ disappears and the left boundary $a\rightarrow 0$.

The large $m$ scaling of the integral becomes manifest after rewriting the effective potential~\eqref{eq:eff-potential} in terms of the density,
\beq
\begin{aligned}
\Gamma &= \frac{2m^2}{\beta} \int\limits_{a}^{b} \rmd x \rho(x)
V\Big(\frac{4\pi m x}{\beta}\Big)
- \frac{4m^2}{\beta^2}  \int\limits_{a}^{b}\int\limits_{a}^{b}  \rmd x \rmd y \rho(x)\rho(y) \log\Big(\frac{16\pi^2 m^2}{\beta^2}|x^2-y^2|\Big) \\
&\approx -2m^2 k \log{\left[\frac{4\pi m}{\beta}\right]} -  \frac{4m^2}{\beta^2} \bigg[ \int\limits_{a}^{b} \rmd x \rho(x) (\log{x}-2\pi x) + \int\limits_{a}^{b}\int\limits_{a}^{b} \rmd x \rmd y \rho(x)\rho(y) \log{|x^2-y^2|}\, \bigg]\, ,
\end{aligned}
\eeq
using eqs.~\eqref{eq:scaledV} and~\eqref{normalisation}. The logarithmic term $\sim m^2\log{m}$ cancels a similar term coming from the overall constant $\C$ in eq.~\eqref{eq:calN}; the other prefactors are subleading at large $m$. This constant can be given in terms of Barnes' $G$-function,
\beq\label{eq:G-barnes}
G(z) \equiv \prod_{i=1}^{z-2} i! \qquad
\Rightarrow \qquad \C = \frac{G(n+m+1)}{G(n-m+1)}\, ,
\eeq
and approximated using the asymptotic behavior,
\beq\label{eq:G-large}
\log{G(z+1)} \approx \frac{z^2}{2}\log{z} -\frac{3z^2}{4} + {\cal O}(z)\, ,
\eeq
for large $z \gg 1 $. Defining then the scaling function
\beq\label{eq:fkdef}
f(k) \equiv \lim_{m, n\rightarrow \infty} m^{-2}\log{\Phi_{m, n}} \approx - \lim_{m, n\rightarrow \infty} m^{-2}(\Gamma+\log{\, \C})\, ,
\eeq
with the limit taken at fixed $k$, we get
\beq\label{eq:fk}
f-\Cp = \frac{4}{\beta^2}\int\limits_{a}^{b} \rmd x \rho(x) (\log{x}-2\pi x) + \frac{4}{\beta^2}\int\limits_{a}^{b}\int\limits_{a}^{b} \rmd x \rmd y \rho(x) \rho(y) \log{|x^2 - y^2|} \, ,
\eeq
where
\beq\label{Cprime}
\Cp = \lim_{m, n\rightarrow \infty} \left( 2k \log{\left[\frac{4\pi m}{\beta}\right]}-m^{-2}\log{\, \C} \right) =  \frac{(k+1)^2}{2} \log{\left[\frac{k-1}{k+1}\right]} + 3k+2k \log{(4\pi)}\, .
\eeq
The saddle-point equation follows from varying $\Gamma = \Gamma[\rho]$, at fixed normalization (\ref{constraint}), giving
\beq \label{eqn-FT-saddlePT}
\lambda = \log{x}-2\pi x + 2\int\limits_{a}^{b} \rmd y \, \rho(y) \log{|x^2 - y^2|} \, ,
\eeq
with $\lambda$ a Lagrange multiplier. A derivative in $x$ eliminates $\lambda$ and yields the singular equation
\beq\label{saddle-point}
0 = \frac{1}{x}-2\pi +\dashint\limits_{a}^{b}\frac{4 x \rho(y)\rmd y}{x^2-y^2}\, , \qquad \forall x\in (a, b)\, ,
\eeq
with $\dashint$ denoting Cauchy principal value.
The equation can also be cast in a more classic form,
\beq\label{eq:sym}
0 = \frac{1}{x}-2\pi \, \sign{\,x} +\dashint\limits_{\mathcal{C}'}\frac{2\rho(y)\rmd y}{x-y}\, ,
\eeq
after unfolding the interval, $(x, y)\in \mathcal{C}' = (-b, -a)\cup (a, b)$, and imposing $\rho(-x) = \rho(x)$.

\begin{figure}[t]
\begin{center}
\includegraphics[scale=0.55]{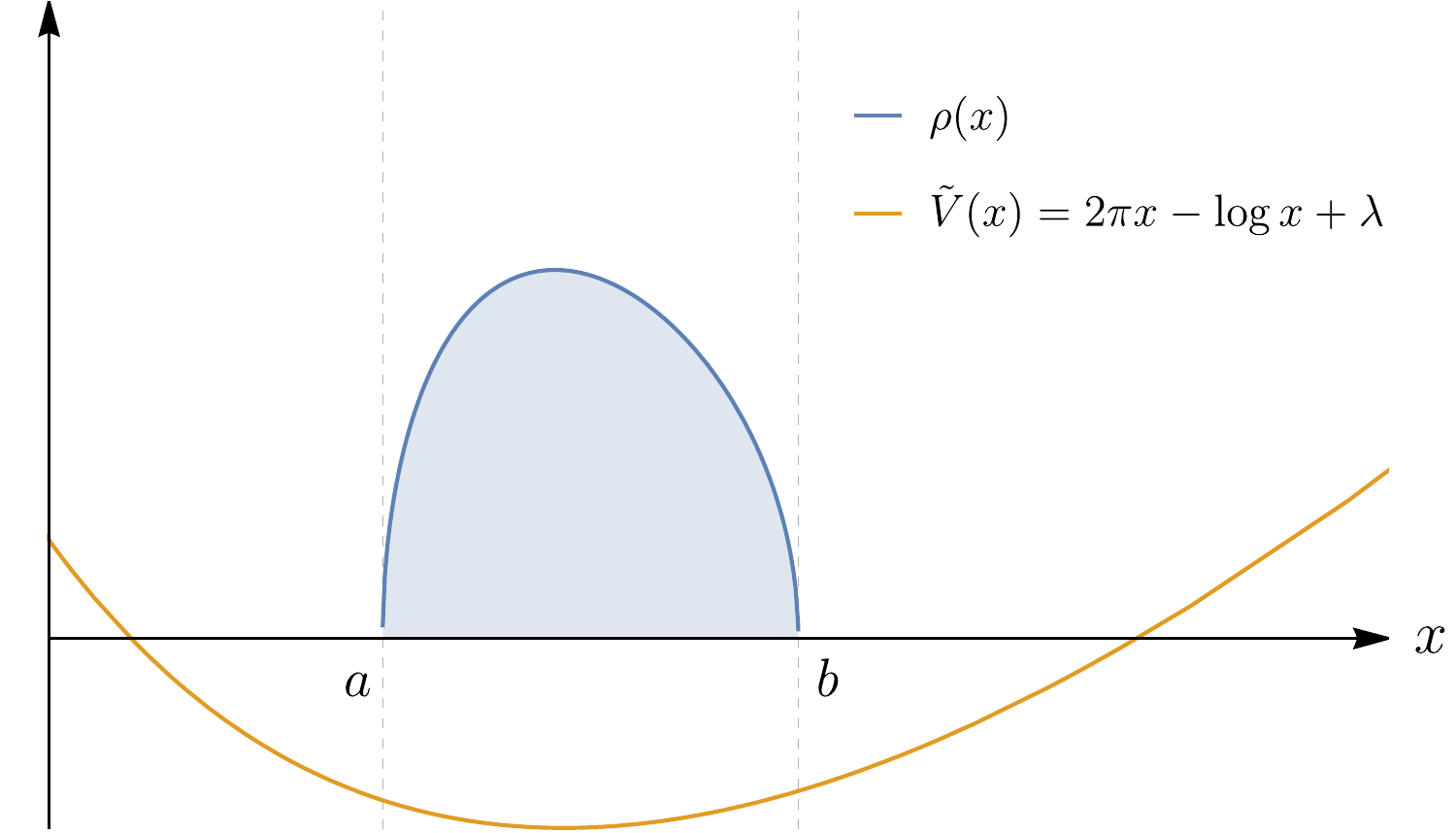}
\end{center}
\caption{Plot of the density $\rho$ and shifted potential $\tilde{V}(x) = 2\pi x -\log{x}+\lambda$ at low density $\beta\equiv 0.015$.} \label{density}
\end{figure}

\subsubsection{Spin-chain mapping}\label{sec:spin-chain}

Equation~\eqref{eq:sym} describes a well-known two-cut matrix-model problem, which appears in various contexts and can be solved exactly (see e.g.~refs.~\cite{DiFrancesco:1993cyw,muskhelishvili2008singular,majumdar2009index,majumdar2011many,grabsch2017truncated,tricomi1985integral} for reviews of general methods). In particular, it was found to control the classical limit of the XXX$_{-1/2}$ Heisenberg spin chain  \cite{Beisert:2003ea, Beisert:2003xu}. In this case, one considers a collection of $S$ magnons propagating on a closed spin chain of length $J$. The magnons' momenta are parametrized by a set of roots $\{u_{j}, j=1, \ldots , S\}$, quantized by the Bethe ansatz equations
\beq
\left(\frac{u_{i}+\I/2}{u_{i}-\I/2}\right)^{J} \prod_{j\neq i}^{S} \frac{u_{i}-u_{j}+\I}{u_{i}-u_{j}-\I} = 1\, , \qquad i = 1, \ldots , S\, .
\eeq
The classical limit corresponds to $J, S \rightarrow \infty$ while keeping the density $S/J$ fixed. In this limit, the roots scale as $u_{i}  = \mathcal{O}(J)$ and the logarithms of the Bethe ansatz equations simplify (see \cite[Appendix C]{Beisert:2003ea}), 
\beq\label{eq:Bethe}
\frac{J}{u_{i}} + \sum_{j\neq i}^{S} \frac{2}{u_{i}-u_{j}} = 2\pi n_{i}\, ,
\eeq
with the so-called mode numbers $n_{i} \in \mathbb{Z}_{\neq 0}$ parametrizing the branches of the logarithms. The ground state corresponds to a symmetric solution, $\{u_j\} = \{-u_j\}$, with the minimal choice, $n_{j} =\sign{\, u_j}$. At large $S$ the roots condense on a symmetric support $\mathcal{C}' = (-b, -a) \cup (a, b)$ and the Bethe equations~\eqref{eq:Bethe} readily reduce to eq.~(\ref{eq:sym}) when written in terms of the root distribution density
\beq\label{eq:rho-spin-chain}
\rho(x) = \frac{1}{J}\sum_{j=1}^{S}\delta (x-u_{j}/J)\, .
\eeq
One also verifies that the normalizations agree if $\beta = S/J$.

Thanks to this identification, one may read off the solution to our problem from the solution given in refs.~\cite{Beisert:2003ea, Beisert:2003xu}. It is expressed in terms of elliptic integrals. In particular, the parameters $a,b$ and $\beta$ are given parametrically as
\be \label{eqn-DiffEqnEllipPara}
a = \frac{1}{4 K(q)}, \qquad b = \frac{1}{4 \sqrt{1-q} K(q)}, \qquad \beta = \frac{E(q)}{2 \sqrt{1-q} K(q)} - \frac{1}{2} \, ,
\ee
with $K(q)$ and $E(q)$ the complete elliptic integrals of the 1st and 2nd kind, respectively,
\beq \label{eqn-defEK}
K(q) = \int\limits_{0}^{\pi/2} \frac{d\phi}{\sqrt{1-q \sin^{2}{\phi}}}\, , \qquad E(q)  = \int\limits_{0}^{\pi/2} d\phi \sqrt{1-q \sin^{2}{\phi}}\, .
\eeq
Here, $q\in (0, 1)$ with the lower/upper bound corresponding respectively to $\beta \rightarrow 0$ and $\beta \rightarrow \infty$. One may also express the density $\rho(x)$ in closed form, in terms of the incomplete elliptic integral of the third kind. (See eqs.~(C.7) and (C.8) in appendix C of ref.~\cite{Beisert:2003ea}.) We plot $\rho(x)$ in figure \ref{density} for illustration. We will not need its explicit analytic expression here, but it will be useful to know its derivative with respect to $\beta$, $\delta \rho = \partial_{\beta}\rho$.

The derivative of the density is rather simple to construct. Differentiating both sides of eq.~\eqref{eqn-FT-saddlePT} with respect to $\beta$, the potential drops out. Using also the vanishing of $\rho$ at the boundaries, $\rho(a) = \rho(b) = 0$, one is left with a simpler (homogeneous) problem
\beq\label{eq:delta-rho}
\partial_{\beta} \lambda = 2\int\limits_{a}^{b}\rmd y \delta \rho(y) \log{|x^2-y^2|} \, , \qquad \forall x \in (a, b)\, .
\eeq
In the Coulomb gas picture, this equation describes a collection of charges with no external potential, but subject to hard-wall boundary conditions at $a$ and $b$. So, despite the absence of a potential, the charges do not run away but accumulate close to the boundaries, $c = a, b$, where $\delta \rho(x) \sim 1/\sqrt{|x - c|}$.

Taking a derivative of eq.~\eqref{eq:delta-rho} with respect to $x$ removes the left-hand side of the equation, and the solution is immediately obtained using a standard inversion formula,
\beq\label{eq:exp-delta-rho}
\delta \rho = \frac{x}{\pi \sqrt{(b^2-x^2)(x^2-a^2)}}\, .
\eeq
The only freedom here is the overall normalization, which is fixed using
\beq
\delta \int \limits_{a}^{b}\rmd x \rho(x) = \int\limits_{a}^{b}\rmd x\,  \delta\rho(x) = \frac{\partial}{\partial \beta} \frac{\beta}{2} = \frac{1}{2} \, ,
\eeq
taking into account again that the boundary terms vanish.

\subsubsection{Differential equation I}

The next step is to calculate the scaling function~\eqref{eq:fk}. This may be done in principle using the exact solution for $\rho$ given in refs.~\cite{Beisert:2003ea, Beisert:2003xu}. Its direct integration proves  difficult however. So, here we shall consider the much simpler problem of determining $df/d\beta$ using the density derivative $\delta \rho = \partial_{\beta}\rho$.

Observe first that eq.~\eqref{eq:fk} simplifies when evaluated on the saddle point. Acting with $\int \rmd x \rho(x)$ on both sides of eq.~(\ref{eqn-FT-saddlePT}) and using the normalisation condition (\ref{normalisation}), one finds the relation
\beq \label{eqn-BMN-BetaLambda}
\tfrac{1}{2}\beta \lambda = \int_{a}^{b}\rmd x \rho(x)(\log{x}-2\pi x) + 2\int_{a}^{b}\int_{a}^{b} \rmd x \rmd y \rho(x)\rho(y) \log{|x^2 - y^2|}  \, ,
\eeq
which may be used to eliminate the double integral in eq.~\eqref{eq:fk},
\beq\label{eq:interm}
\beta^2 (f-\Cp) = \beta \lambda + 2 \int_{a}^{b}\rmd x \rho(x)(\log{x}-2\pi x) \, ,
\eeq
with $\Cp$ given in eq.~\eqref{Cprime}. A derivative with respect to $\beta$ then yields
\beq\label{eq:firstly}
\frac{d}{d\beta}(\beta^2 (f-\Cp)) = \lambda+ \beta\partial_{\beta}\lambda + 2 \int_{a}^{b}\rmd x \, \delta\rho(x)(\log{x}-2\pi x) \, ,
\eeq
using that $\delta \int = \int \delta$ since $\rho$ vanishes at the boundary.

One may further simplify the right-hand side of eq.~\eqref{eq:firstly}, by acting on eq.~\eqref{eqn-FT-saddlePT} and eq.~\eqref{eq:delta-rho} with $2\int \rmd x \delta \rho(x)$ and $2\int \rmd x \rho(x)$, respectively, and taking their difference,
\beq
\lambda = \beta\partial_{\beta}\lambda + 2\int_{a}^{b}\rmd x \, \delta\rho(x)(\log{x}-2\pi x) \qquad \Rightarrow \qquad \frac{d}{d\beta}(\beta^2 (f-\Cp)) = 2\lambda\, .
\eeq
The integral on the right-hand side of the equation for $\lambda$, as well as the one defining $\partial_{\beta}\lambda$, see eq.~\eqref{eq:delta-rho}, can be taken immediately using eq.~\eqref{eq:exp-delta-rho},
\begin{align}
&\partial_{\beta} \lambda = \log{\left[\frac{b^2 - a^2}{4}\right]}\, , \qquad \int_{a}^{b}\rmd x \, \delta\rho(x)(\log{x}-2\pi x) = \frac{1}{2}\log{\left[\frac{a+b}{2}\right]} -\beta-\frac{1}{2}\, .
\end{align}
Finally, eliminating $\lambda$ yields
\beq \label{eqn-DiffEqn01}
\frac{d}{d\beta}\Big[ \beta^2 (f-\Cp) \Big]
= 2\beta \log{\left[\frac{b^2 - a^2}{4}\right]} + 2\log{\left[\frac{a+b}{2}\right]} -4\beta-2\, ,
\eeq
with $a, b, \beta$ related to each other through eq.~\eqref{eqn-DiffEqnEllipPara}. 

This first-order differential equation determines $f$ uniquely, once supplemented with the boundary condition $\lim_{\beta\rightarrow 0}(\beta^2 f) = 0$. The boundary condition follows from~\eqref{eq:interm}, using that $b\rightarrow a$ when $\beta \rightarrow 0$, and $\lim_{\beta\rightarrow 0}(\beta^2 \Cp) = 0$ from eq.~\eqref{Cprime}.

The differential equation~\eqref{eqn-DiffEqn01} can be integrated directly around particular points, such as $\beta =0$ or $\infty$, after evaluating its right-hand side using eqs.~\eqref{eqn-DiffEqnEllipPara}. It is less straightforward to integrate it for general $\beta$, given the complicated dependence of the coefficients $a, b$ on $\beta$. Fortunately, we will not need to deal with this problem here, as the solution will come for free after combining eq.~\eqref{eqn-DiffEqn01} with a similar equation controlling the thermodynamic limit of the BMN integral.

\subsection{BMN integral}

Consider now the BMN integral~\eqref{eq:I_BMN}. In this case, instead of a single matrix-model integral, we have an ensemble of integrals, labelled by integers $\{a_\ell\}$. However, at large $m$, we expect it to be dominated by the lowest modes, with $a_{\ell} = 1, \forall \ell=1, \ldots , m$. This is because the external potential
\beq\label{eq:Va}
V_{a}(u) = (m+n) \log{(u^2+a^2/4)}
\eeq
has its minimum at $a=1$ and generates power suppressions for the higher $a$'s in the thermodynamic limit. Owing to this truncation, there is no dependence on $\sigmap$, and similarly $\sigmam$ drops out, for $\sigma, \varphi = \mathcal{O}(1)$. In summary, focusing on the leading saddle, one may write
\beq\label{BMN-continuum}
\Phi_{m, n} \approx \frac{d(z, \bar{z})^{m}}{(2\pi)^m m!}\times\int_{\mathbb{R}^m} \prod_{\ell=1}^{m}\frac{\rmd u_{\ell}}{(u_{\ell}^2+1/4)^{m+n}} \prod_{i<j}^{m} \left[(u_{i}-u_{j})^2((u_{i}-u_{j})^2 + 1)\right]\, .
\eeq
Furthermore, one notes that the potential~\eqref{eq:Va} scales linearly with $m$, for all $u$'s. Hence, there is no need to rescale the rapidities here and we can proceed directly to the thermodynamic limit with the density
\beq
\rho(u) = \frac{1}{m}\sum_{i=1}^{m}\delta(u-u_i)\, ,
\eeq
which we expect to be smooth for $u= \mathcal{O}(m^0)$. Varying the integral~\eqref{BMN-continuum}, we get
\beq \label{eqn-BMN-saddle}
0 = \frac{(k+1) u}{u^2+1/4} - \dashint_{-B}^B \rmd v \rho(v) \Big[ \frac{u-v}{(u-v)^2+1} + \frac{1}{u-v} \Big]\, , \qquad \forall u \in (-B, B)\,,
\eeq
assuming that the density is even and supported on a compact interval $(-B, B)$, as suggested by the symmetry and convexity of the potential. We seek a solution obeying the usual boundary conditions
\beq\label{BC-rho}
\rho(u) \propto \sqrt{B^2-u^2}\, , \qquad u\sim \pm B\, , 
\eeq
and normalized such that
\beq
\int_{-B}^B \rmd v \, \rho(v) = 1 \, .
\eeq

\subsubsection{Resolvent}

Equation~(\ref{eqn-BMN-saddle}) can be solved immediately in the two extreme regimes, $k\rightarrow \infty$ and $k\rightarrow 1$. The former corresponds to a dilute gas approximation, $B\rightarrow 0$, where the singular part $\sim 1/(u-v)$ of the kernel dominates and the potential becomes linear. Thus it reduces to the Gaussian matrix model and the associated semi-circle law
\beq
\rho(u) \approx \frac{2}{\pi B^2} \sqrt{B^2-u^2}\, ,
\eeq
with $B\sim 1/\sqrt{2k}$. The opposite limit, $B\rightarrow \infty$, maps to $k \rightarrow 1$, and the equation may be solved by a Fourier transform over the entire real axis,
\beq\label{eq:large-B}
\lim_{k\rightarrow 1}\rho(u) = \sech{(\pi u)}\, .
\eeq
The problem gets harder in the intermediate regime $k\in (1, \infty)$. It can nonetheless be solved exactly by following a method introduced in ref.~\cite{Kazakov:1998ji} to tackle a similar looking matrix-model equation.

\begin{figure}[t]
\centering
\includegraphics[width=0.3\textwidth]{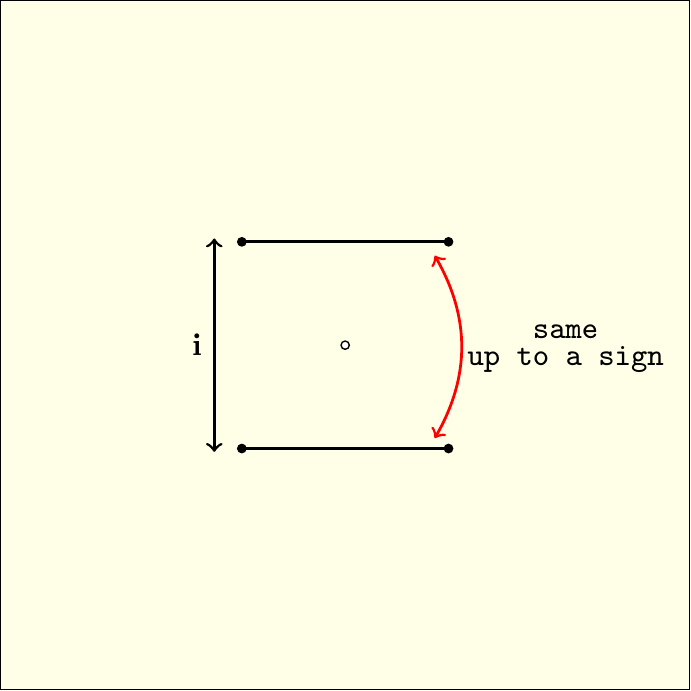}
\caption{Riemann sheet structure of the resolvent $\g(u)$. There is a simple pole at the origin (and at infinity) and two cuts of length $2B$ spaced by $\I$. The periodicity condition identifies the values of the functions above and below the cuts, up to a sign.}
\label{cut-planes}
\end{figure}

To begin, we define the function
\be
\g(u) = \frac{k+1}{u} - \int\limits_{-B}^{B} \rmd v  \frac{2(u-v) \rho(v)}{(u-v)^2+1/4}\, .
\ee
It is analytic in $\mathbb{C}$, except along the intervals $(- B \pm \I/2, + B \pm \I/2)$, where it has square-root branch cuts, and at $u = 0$, where it has a simple pole, $r(u)\sim (k+1)/u$. We also note that the function has a pole at $u = \infty$, with residue
\beq
\lim_{u\rightarrow \infty} u\,\g(u) = k+1 -2 \int\limits_{-B}^{B} \rmd v \rho(v) = k-1\, .
\eeq
The function $\g$ relates to the standard (single-cut) resolvent
\beq
R(u) = \int\limits_{-B}^{B} \frac{\rmd v \rho(v)}{u-v} \qquad \Rightarrow \qquad \g(u) = \frac{k+1}{u} - (R(u+\I/2)+R(u-\I/2))\, ,
\eeq
implying that one may recover the density $\rho(u)$ from the discontinuities of $\g(u)$ across either cut, 
\beq
\rho(u) = \frac{1}{2\pi \I} \Bigl[ \g(u\pm \I/2 +\I 0)-\g(u\pm \I/2 -\I 0) \Bigr]\, , \qquad \forall u\in (-B, B)\, .
\eeq
In particular, it follows from this relation that $r(u)$ must be continuous at the branch points $u = \pm \I/2 \pm B$, since the density vanishes at $u=\pm B$, see eq.~\eqref{BC-rho}.

The main motivation for introducing a resolvent in this way is that the saddle-point equation~\eqref{eqn-BMN-saddle} now takes the simple form
\be\label{disc}
\g(u+\I/2-\I 0) + \g(u-\I/2+\I0) = 0, \qquad \forall u \in (-B,B) \, .
\ee
It implies that $\g(u)$ extends to an anti-periodic function of period $\I$ on the second Riemann sheet, that is, after continuing below the cuts, see figure \ref{cut-planes}.

\begin{figure}[t]
\centering
\begin{minipage}[c]{0.33\linewidth}
\centering
\includegraphics[width=0.7\textwidth]{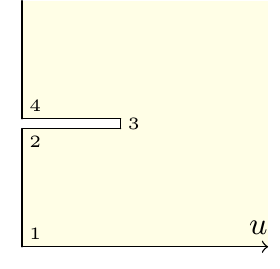}
\end{minipage}
\hspace*{-1cm}
$\xlongleftrightarrow{\textrm{SC}}$
\hspace*{-1cm}
\begin{minipage}[c]{0.33\linewidth}
\centering
\includegraphics[width=0.7\textwidth]{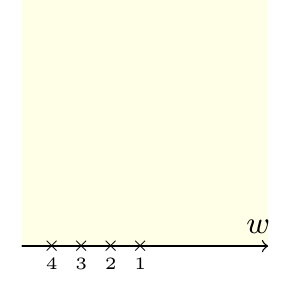}
\end{minipage}
\begin{minipage}[c]{0.33\linewidth}
\centering
\begin{tabular}{|c|c|} \hline
\,\, $u$  \,\, & \, $w$ \, \\ \hline
\,\, $u_{1} = 0$ \,\,  & \,$w_1 = 0$ \, \\ 
\,\, $u_{2} = \frac{\I}{2} - \I 0$ \,\,  & \, $w_2$ \, \\ 
\,\, $u_{3} = \frac{\I}{2} + B$ \,\, & \,  $w_3$ \, \\ 
\,\, $u_{4} = \frac{\I}{2} + \I 0$ \,\, & \, $w_4$ \, \\ 
\,\, $\infty$ \,\, & $\infty$ \, \\  \hline
\end{tabular}
\end{minipage}
\caption{The conformal mapping $u \rightarrow w = 1/\g(u)^2$ maps the domain $D$ to the upper half-plane $\UHP$, and vice versa. In particular, the boundaries map into each other owing to the reflection symmetry and periodicity of $\g(u)$. The table indicates the values of the map at the marked points on the boundary.
}\label{fig-ConfMapping}
\end{figure}

In addition, the resolvent $\g$ possesses important reality and reflection properties,
\be\label{eq-reality}
\g(u) =  (\g(u^*))^*\, , \qquad \g(u) = - \g(-u) \, ,
\ee
with $*$ denoting complex conjugation, which stem from the fact that $\rho$ is a real symmetric function. It then suffices to find $\g$ on the first quadrant, in order to determine it everywhere else. Taking the cuts into account, we should in fact consider the open domain $D$ shown in the left panel of figure \ref{fig-ConfMapping}, which is obtained by removing the segment between $u = \I/2$ and $u = B+ \I/2$ from the first quadrant.

\subsubsection{Schwarz-Christoffel mapping}

The function $r$ obeys simple functional relations, but is defined over a complicated domain. The next step is to make the domain more regular with the help of a Schwarz-Christoffel (SC) transformation. Namely, $D$ can be seen as the interior of a polygon (with a vertex at infinity) and as such can be mapped conformally to the upper half-plane $\UHP = \{w\in \mathbb{C}: \textrm{Im}\, w > 0\}$ with the boundary $\partial D$ mapping to the real $w$ axis. The map is constructed canonically, using the interior angles at the vertices of $\partial D$ to determine the exponents in the SC transform $w\rightarrow u$. Fixing $u = w = 0$ and $u = w = \infty$ for simplicity, we get
\be \label{eqn-Ellip-z}
u  = A \int_{0}^w \frac{(t - w_3)\rmd t }{\sqrt{t(t-w_2)(t-w_4)}}\, ,
\ee
with $w_4 < w_3 < w_2 < 0$ the images of $u_{4} =  \I/2+\I0\, , u_{3} = \I/2+B$ and $u_{2} = \I/2-\I0$, respectively, and with $A> 0$ an arbitrary normalization factor, introduced for later convenience.

We may now easily express the resolvent $r$ as a function of $w$. The key observation is that the map
\beq
\mathcal{F} : w \rightarrow \mathcal{F}(w) = 1/\g(u(w))^2
\eeq
is a holomorphic bijective function of $\UHP$ to itself, that is, a real M\"obius transformation. This is manifest in the neighbourhood of $w = 0$ or $w=\infty$. At these points, $\g(u)\sim (k\pm 1)/u$ and $\sqrt{w}/u$ is constant. Hence, one can always write
\beq\label{eqn-Ellip-gz}
r(u) = \frac{1}{\sqrt{w}}\, , 
\ee
for a suitable choice of the parameters $A, w_{2,3,4}$, see eq.~\eqref{eqn-Ellip-Final} below. The claim is that eq.~\eqref{eqn-Ellip-gz} holds globally, for all $w\in \UHP$.

To justify the claim, it is enough to show that $\mathcal{F} $ extends to a continuous function of the closed upper half-plane taking real values on the real axis. The Schwarz reflection principle can then be used to conclude that $\mathcal{F}$ extends to an entire function on $\mathbb{C}$%
\footnote{The extension is given by $\mathcal{F} (w^*) = \mathcal{F}(w)^*$ and coincides with the continuation of the SC map~\eqref{eqn-Ellip-z} to the domain $u\in D \cup D^*$.} and thus $\mathcal{F}(w) = w$ given the asymptotic behaviours at $0$ and $\infty$. 

Now, it is easy to show that $\mathcal{F}$ preserves the boundary $\partial \UHP$. This follows directly from the reality and periodicity properties of $\g$. Namely, $\g(u)$ is respectively real for $u\in (0, \infty)$ and purely imaginary for $u\in (0, \I \infty)$, because of eq.~\eqref{eq-reality}. It is also purely imaginary just below the cut, i.e.~$u\in (u_{2}, u_{3})$, owing to eq.~\eqref{disc},
\beq
\g(u+\I/2 - \I 0)^* = \g(u-\I/2 + \I 0) = -\g(u+\I/2 - \I 0) \, ,
\eeq
and similarly for $u\in (u_{3}, u_{4})$, thanks to the discontinuity formula \eqref{disc} and the reality of $\rho$.%
\footnote{Alternatively, one notices that $\g(u+\I/2+\I0) + \g(u-\I/2-\I0) = 0$ after combining eqs.~\eqref{eq-reality} and \eqref{disc}.} Furthermore, $\mathcal{F}$ is continuous on $\partial \UHP$, including at the fold-point $w = w_3$, since $\rho(u)$ vanishes at $u = u_3$, and at the origin $w=0$, where $\mathcal{F} \rightarrow 0$. Hence, as desired, $\mathcal{F}$ maps $\partial \UHP$ into itself.

Finally, one may determine the unfixed parameters $(A, w_{2,3,4})$ in the SC transformation, by matching both sides of eq.~\eqref{eqn-Ellip-gz} at the marked points on the boundary. The analysis is deferred to appendix~\ref{sec:parameter}. The results can be written concisely in terms of the elliptic data $a, b , q$ introduced in eqs.~\eqref{eqn-DiffEqnEllipPara},
\bea \label{eqn-Ellip-Final}
A  = \frac{1}{2 \beta}\, ,\qquad w_2  = \frac{-\beta^2 }{16 \pi^2 b^2}\, , \qquad w_3  = \frac{-\beta^2 (1+2\beta) }{16 \pi^2 a b}\, , \qquad w_4  = \frac{-\beta^2 }{16 \pi^2 a^2}\, ,
\eea
with $\beta = 1/(k-1)$ and with the modulus $q = 1-w_2/w_4 \in (0, 1)$.

\subsubsection{Duality map}

Prior to moving on to the calculation of the free energy, let us open a parenthesis here to comment on the relation between the above solution and the one constructed earlier. The two solutions should be images of each other, as they are based on two equivalent representations, for any $m$ and $n$. In practice, however, it is not easy to follow the Mellin-Barnes transformation used in section~\ref{sec:BMN_det} all the way to the large $m$ limit. One may nonetheless find the duality map directly in this limit from a comparison of the two saddle-point solutions.

Recall first of all that the interval $(a, b)$ determines the support of the dual density, denoted $\rho_{\textrm{dual}}(x)$ here, to avoid confusion. Looking at~\eqref{eqn-Ellip-Final}, we see that this interval can be identified with the segment $(w_{2}, w_{3})$ of the $w$ plane. More precisely, the above relations suggest to set
\beq\label{eq:x-r}
x = \frac{\I \beta}{4\pi \sqrt{w}} = \frac{\I \beta r(u)}{4\pi}\, ,
\eeq
thus identifying the image of the resolvent $r(u)$ with the dual $x$ plane. The reciprocal relation identifies the $u$ plane with the image of a suitable resolvent $r_{\textrm{dual}}(x)$ for the dual problem,
\beq
u = \frac{\I}{2\pi} r_{\textrm{dual}}(x)\, .
\eeq
This relation can be worked out from the elliptic parametrization~\eqref{eqn-Ellip-z} together with~\eqref{eq:x-r}. One may also proceed as follows. Using the map \eqref{eq:x-r} and properties of $r(u)$, one easily sees that the inverse map $r_{\textrm{dual}}(x) = -r_{\textrm{dual}}(-x)$ is an analytic function of $x$ with square-root cuts along $(-b, -a)\cup (a, b)$ and poles at $x=0$ and $x=\infty$, with residues
\beq\label{eq:res-rdual}
r_{\textrm{dual}}(x) \sim \frac{1}{2x}\, , \qquad r_{\textrm{dual}}(x) \sim \frac{k+1}{2 (k-1) x}\, ,
\eeq
respectively. It must also obey the equation
\beq\label{eq:sum-rdual}
r_{\textrm{dual}}(x+\I 0) + r_{\textrm{dual}}(x-\I 0) = 2\pi \I (u^*-u) = 2\pi\, ,
\eeq
for any $x\in (a, b)$, since on this interval $\textrm{Im}\, u = \I/2$ according to eq.~\eqref{eq:x-r}, see table in figure~\ref{fig-ConfMapping}. These requirements define a Riemann-Hilbert problem whose solution is given by
\beq
r_{\textrm{dual}}(x) = \frac{1}{2x} + \int\limits_{a}^{b} \frac{2x \rho_{\textrm{dual}}(y)\rmd y}{x^2-y^2}\, ,
\eeq
for $x\notin (a, b)$, with $\rho_{\textrm{dual}}(x)$ the density constructed in the previous section. In particular, equations~\eqref{eq:res-rdual} and~\eqref{eq:sum-rdual} are easily checked using~\eqref{normalisation} and the dual saddle-point equation~\eqref{saddle-point}. At last, $r_{\textrm{dual}}(x)$ is seen to be a resolvent for the dual problem, since
\beq
\rho_{\textrm{dual}}(x)  = \frac{\I}{2\pi}(r_{\textrm{dual}}(x+\I 0) - r_{\textrm{dual}}(x-\I 0))  = \textrm{Re}\, u\, ,
\eeq
for any $x\in (a, b)$. Hence, nicely, the large $m$ duality map merely interchanges rapidities and resolvents of the two problems.

\subsubsection{Differential equation II}

We may now calculate $df/dk$ using the density of the BMN integral (\ref{BMN-continuum}). The starting point is the saddle-point expression obtained from eq.~\eqref{BMN-continuum},
\be
f(k) = - (1+k) \int \rmd u \rho(u) \log (u^2 + \tfrac{1}{4}) + \frac{1}{2} \int \rmd u \rmd v \rho(u) \rho(v) \log [(u-v)^2 ((u-v)^2+\tfrac{1}{4})] \, ,
\ee
which is extremal on the just-constructed solution $\rho$,
\be\label{eq:mu}
\mu = \frac{\delta f}{\delta \rho(u)} =  - (1+k)\log (u^2 + \tfrac{1}{4}) + \int \rmd v \rho(v) \log [(u-v)^2 ((u-v)^2+\tfrac{1}{4})]  \, ,
\ee
with $\mu$ a Lagrange multiplier for $\int \rho(u) \rmd u = 1$. These relations imply that
\bea\label{eq-dfdk}
\frac{\rmd f}{\rmd k} & =  - \int \rmd u [\rho(u)+(k+1)\delta \rho(u)] \log (u^2 + \tfrac{1}{4}) +  \int \rmd u \rmd v  \delta \rho(u) \rho(v) \log [(u-v)^2 ((u-v)^2+\tfrac{1}{4})] \\
& = - \int \rmd u \rho(u) \log (u^2 + \tfrac{1}{4}) + \mu \int \rmd u \, \delta \rho(u) = - \int \rmd u \rho(u) \log (u^2 + \tfrac{1}{4}) \, ,
\eea 
with $\delta \rho = d\rho/dk$, after eliminating the double integration with eq.~\eqref{eq:mu} and using $\delta \int \rho = \int \delta \rho = 0$.

We then introduce an auxiliary function,
\be
h(u) = \int_{-B}^{B}\rmd v \rho(v) \log [(u-v)^2 + \tfrac{1}{4}] \, ,
\ee
which returns $-df/dk$ when evaluated at $u = 0$. This function is nothing but an integrated version of the resolvent $r(u)$, with given large $u$ behavior,
\bea\label{eqn-FunAux}
\frac{\rmd h(u)}{\rmd u} & =  \frac{k+1}{u}-\g(u) \, , \qquad h(u) = 2 \log{u} + \mathcal{O}(1/u^2)\, .
\eea
This equation can be integrated with the help of eqs.~\eqref{eqn-Ellip-z} and~\eqref{eqn-Ellip-gz}. It yields
\be
\begin{aligned}
h(u) &= (k+1)\log{u} -(k-1)\int^{w} \frac{\rmd t}{2t} \frac{t-w_3}{\sqrt{(t-w_2)(t-w_4)}} \\
&= (k+1) \log{u} - (k-1) \log{(\sqrt{w-w_4}+\sqrt{w-w_2})} + (k+1) \, \arctanh{\, \sqrt{\frac{(w-w_4)w_2}{(w-w_2)w_4}}} +cte \, , \\
\end{aligned}
\ee
up to a constant of integration, fixed by the large $u$ asymptotics,
\beq
cte = -(k-1)\log\Big[ \tfrac{1}{2}(k-1) \Big] - (k+1)\log{\sqrt{\frac{b+a}{b-a}}}\, ,
\eeq
using $u\sim (k-1)\sqrt{w} \gg 1$ and eq.~(\ref{eqn-Ellip-Final}). Lastly, the limit $u \ra 0$ yields
\be \label{eqn-DiffEqn02}
\frac{\rmd f}{\rmd k} =2k \log\Big[ \tfrac{1}{2}(a+b) \Big] - (k-1) \log{(a b)} + 2\log{(4\pi)} + (k+1) \log{\left[\frac{k-1}{k+1}\right]} \, ,
\ee
using $u \sim (1+k)\sqrt{w} \ll 1$ (see appendix \ref{sec:parameter}) and switching to variables $a,b$ with eq.~\eqref{eqn-Ellip-Final}.

\subsection{Free energy}

We shall now derive a parametric representation for the scaling function $F(k) = f(k)/k$.  From eq.~\eqref{eq:fkdef} we see that $F(k)$ is normalized by the rectangle area $mn$ rather than $m^2$.

\subsubsection{Parametric representation}

A key simplification arises when combining our two differential equations, eqs.~\eqref{eqn-DiffEqn01} and \eqref{eqn-DiffEqn02}. Namely, one finds that the function $f$ can be extracted directly, without any integration. This is possible because the equations are linearly independent and have only one solution in common. Taking the sum of the two equations eliminates the derivatives of $f$,
\beq\label{eq-2-diff}
\frac{d}{d\beta}(\beta^2(f-\Cp)) + \frac{d}{d k} f = 2\beta f- \frac{d}{d\beta}(\beta^2\Cp)  \, ,
\eeq
giving%
\footnote{We assume here that the two differential equations are set in the \textit{same} parametrization, that is, that the moduli $q$ are the same. This is the case in the range $\beta \in (0, \infty) \leftrightarrow q\in (0, 1)$, where the elliptic parametrization is bijective,
\be
\frac{d \beta}{d q} = \frac{(K(q)-E(q)) (E(q)-(1-q) K(q))}{4 (1-q)^{3/2} q K(q)^2} > 0, \qquad \forall q\in(0,1) \, ,
\ee
with the lower bound being observed in the limit $q \rightarrow 0$ and with $\beta$ ranging between $0$ and $\infty$ for $q \in (0,1)$. Hence, all values of $k = 1+1/\beta \in (1, \infty)$ are covered once and only once.}
\beq
f = \log\Big[ \tfrac{1}{2}(b-a) \Big] + k^2\log\Big[ \tfrac{1}{2}(a+b) \Big] - \frac{(k-1)^2}{2}\log{(ab)} +2k\log{(4\pi)} +\frac{(k+1)^2}{2} \log{\left[\frac{k-1}{k+1}\right]}\, ,
\eeq
after using eqs.~\eqref{eqn-DiffEqn01},~\eqref{eqn-DiffEqn02},~\eqref{Cprime} and $\beta = 1/(k-1)$.

As a double check, one may verify that this expression solves each differential equation separately.  For this purpose, it is useful to have the $\beta$ derivatives of $a$ and $b$,
\be\label{ab_beta_deriv}
\frac{da}{d\beta} = \frac{2a^2}{2a\beta+a-b} \,,
\qquad \frac{db}{d\beta} = \frac{2b^2}{2b\beta+b-a} \,,
\ee
which follow from eq.~\eqref{eqn-DiffEqnEllipPara} and the chain rule.

Introducing $F(k) = f(k)/k$ and using eq.~\eqref{eqn-DiffEqnEllipPara}, one arrives at
\beq\label{main-F}
\begin{aligned}
F(k) = \log{\pi^2} &+ k\log\Big(\frac{1+\sqrt{1-q}}{2}\Big)
+ \frac{(k-1)^2}{2k}\log{K(q)} \\
& + \frac{1}{k}\log\Big(\frac{1-\sqrt{1-q}}{2}\Big) -\frac{(k+1)^2}{2k}\log{E(q)}\, , \\
\end{aligned}\
\eeq
where we recall that
\beq\label{eq-k-bis}
k = \frac{E(q)+\sqrt{1-q}K(q)}{E(q)-\sqrt{1-q}K(q)}\, ,
\eeq
with $K(q)$ and $E(q)$ the complete elliptic integrals, see eq.~\eqref{eqn-defEK}, and $q \in (0, 1)$.

\begin{figure}[t]
\begin{center}
\includegraphics[width=0.65 \textwidth]{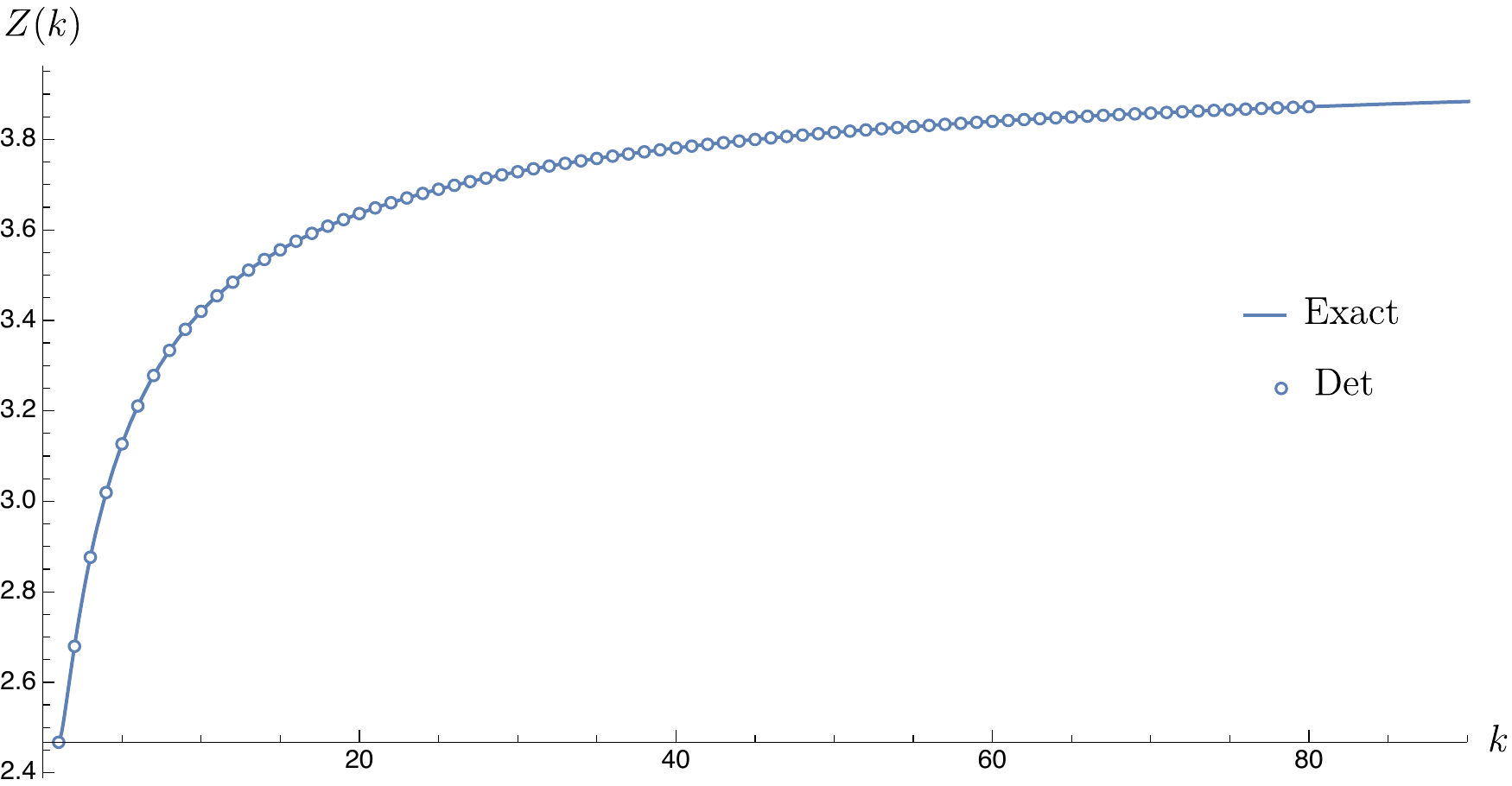}
\end{center}
\caption{Plot of $Z(k) = e^{F(k)}$ for $k\in [1, 90]$. The data points are obtained by extrapolating numerically the determinant at large $m, n$ for $z = \bar{z} = -1$ and $k \in \{1, 80\}$. The solid line is obtained using the parametric representation, eqs.~\eqref{main-F} and~\eqref{eq-k-bis}. $Z(k)$ is seen to be monotonic and bounded by $Z(1) = \pi^2/4 = 2.4674...$ and $Z(\infty) = 4$.}\label{fig-zzbarInd}
\end{figure}

The system of equations~\eqref{main-F} and \eqref{eq-k-bis} is our final expression for the free-energy density. It determines $F(k)$ parametrically over the whole domain $k \in (1, \infty)$. Figure \ref{fig-zzbarInd} shows the comparison between this scaling function and a numerical estimate of the ladder determinant~\eqref{eq:I-det} for large $m$ and $n$. As one can see, the agreement is excellent for all accessible values of $k$.

Note that one may first simplify the determinant to facilitate the comparison. Since the cross ratios drop out in this regime, one can conveniently set $\sigmap = \sigmam = 0$, or equivalently send $z, \bar{z} \rightarrow -1$, in eq.~\eqref{eq:Phi-I}, giving
\beq
\Phi_{m, n} \approx \frac{1}{\mathcal{N}} \det\, (a_{i+j+n-m-1})_{1\leqslant i, j \leqslant m}\, ,
\eeq
with $a_{p} = 8(1-4^{1-p})(2p-1)!\zeta(2p-1)$, using $L_{p}(z, \bar{z}) \approx 2(z-\bar{z}) (1-4^{1-p}){{2p-1}\choose{p-1}} \zeta(2p-1)$.

Below, we study in detail the behaviour of the free energy at $k = \infty$ and $k = 1$, using known asymptotic expressions for the elliptic integrals at $q = 0$ and $q=1$. These points are of particular interest since they correspond to the ladder limit $n\gg m$ and to the symmetric point $n = m$, respectively.

\subsubsection{Ladder limit}

The limit $k\rightarrow \infty$ corresponds to $q\rightarrow 0^+$. The elliptic integrals are regular at this point and admit convergent expansions in integer powers of $q$,
\beq
K(q) = \frac{\pi}{2} + \frac{\pi q}{8} + \mathcal{O}(q^2) , \qquad E(q) = \frac{\pi}{2} - \frac{\pi q}{8} + \mathcal{O}(q^2) \, ,
\eeq
from which it follows that
\beq
q = \frac{4\sqrt{2}}{\sqrt{k}} - \frac{16}{k}+ \mathcal{O}(\frac{1}{k^{3/2}})\, ,
\eeq
using eq.~\eqref{eq-k-bis}. Expanding further and plugging the series inside eq.~\eqref{main-F}, one finds
\beq\label{small-k}
F(k) = \log{4} -\frac{2\log{\left(\frac{k}{2}\right)}+3}{4k} -\frac{1}{8k^2} + \frac{19}{384 k^3}-\frac{7}{512 k^4} +\frac{301}{30720 k^5} -\frac{81}{20480 k^6} + \mathcal{O}\Big(\frac{1}{k^7}\Big)\, .
\eeq
Notice that the half-integer powers of $1/k$ in the expansion of $q$ cancel out in $F$. Hence, $F$ admits a regular expansion in $1/k$, up to a logarithm $\sim \log{k}/k$. The series is seen to alternate and it approximates the exact curve extremely well throughout the entire domain, including the point $k=1$, where \eqref{small-k} is found to converge rather quickly towards the exact value $F(1) = \log{(\pi^2/4)}$. It suggests that the series converges for all $k\in [1, \infty)$.
 
The leading term in this expansion has a simple interpretation. As seen earlier, the limit $k\rightarrow \infty$ corresponds to a dilute gas approximation, which means that the free energy is a multiple of the free energy of an individual ladder $L_{p}$ at large $p$. This is easily verified using the definition~\eqref{ladders}. At large weight, the polylogarithm series~\eqref{ladders} truncates,
\beq\label{eq:large-L}
\textrm{Li}_{j}(z) \approx z \, ,
\eeq
assuming $|z|<1$ for convenience, and the sum $\sum_{j = p}^{2p}$ in eq.~\eqref{ladders} is dominated by the terms with $j \sim 2p$. One can thus take the large $p$ limit of the summand at fixed $l = 2p-j$, using Stirling's approximation for the factorials, and perform the sum over  $l = 0, 1, \ldots\, $,
\beq
L_{p}(z, \bar{z}) \approx (z-\bar{z})  \sum_{l=0}^{p} \frac{(2p-l)!}{p!\, l!\, (p-l)!} [-\log{(z\bar{z})}]^{l} \approx  \frac{4^p (z-\bar{z})}{\sqrt{\pi p z\bar{z}}} \, .
\eeq
It yields $\lim_{p \rightarrow \infty} \frac{1}{p} \log{\left[L_{p}(z, \bar{z})\right]} = \log{4} = 1.386...\, ,$ in agreement with $F(\infty)$.

\subsubsection{Square limit}

The limit $k\rightarrow 1$ corresponds to $q\rightarrow 1^-$ and in this limit
\beq
\lim\limits_{k\rightarrow 1}F(k) = \log{(\pi^2/4)} = 0.903...\, .
\eeq
Subleading corrections have a complicated pattern, involving not only powers of $(k-1)$ but also of its logarithms. This feature goes back to logarithms in the expansion of the elliptic integrals around $q = 1$, 
\beq
\begin{aligned}
&K(q) = -\frac{1}{\pi}K(1-q) \log{(1-q)} + \ldots\, ,\\
&E(q) = -\frac{1}{\pi} (K(1-q)-E(1-q))\log{(1-q)} + \ldots\, ,\\
\end{aligned}
\eeq
with the dots standing for regular series in $(1 - q)$.

A double expansion is easily generated as follows. First, define the variable
\beq\label{eq:zeta}
\zeta = \frac{k-1}{\sqrt{k}} = \frac{2\sqrt{1-q}\, K(q)}{\sqrt{E(q)^2-(1-q)K(q)^2}}\, ,
\eeq
which is small when $k\rightarrow 1$ and changes sign under $k\rightarrow 1/k$, and then introduce the ansatz
\beq\label{eq:ansatz}
q = 1- 16\zeta^2 \bigg[ \F^2(\log \zeta) + \sum_{n=1}^\infty \zeta^{n} \F_n(\log \zeta) \bigg] \, ,
\eeq
with the expansion coefficients, $\F$ and $\F_n$, being functions of $\log{\zeta}$. The latter coefficients are determined self-consistently, by plugging \eqref{eq:ansatz} inside eq.~\eqref{eq:zeta}, expanding in powers of $\zeta$ and matching the coefficients on both sides of the equation. For instance, at the leading order in $\zeta$, one finds
\beq\label{eq:match}
\zeta = -8  \zeta \F \log{(\zeta \F)} + \mathcal{O}(\zeta^2) \, ,
\eeq
or equivalently
\beq\label{eq:alpha}
\alpha  = -\frac{1}{8\log{\zeta}}\left[1+\frac{\log{(-8\log{\zeta})}}{\log{\zeta}} + \mathcal{O}\left(\frac{\log^{2}{(-\log{\zeta})}}{\log^2{\zeta}}\right)\right]\, .
\eeq
The full solution for $\alpha = \alpha(\log{\zeta})$ can also be given in closed form,
\beq
\F = - \frac{1}{8 W_{-1}(-\frac{\zeta}{8})}\, ,
\eeq
with $W_{-1}(x)$ the second branch of the Lambert $W$ function, defined for $x \in (-1/e, 0)$ and such that
\beq
W_{-1}(x) = \log{(-x)} - \log{(-\log{(-x)})} +\ldots\, ,
\eeq
when $x\rightarrow 0^-$.

Cancelling the terms at higher orders in eq.~\eqref{eq:match} determines the higher coefficients $\alpha_{n}$ recursively. One finds that $\alpha_{n} = 0$ when $n$ is odd, in line with the $k\leftrightarrow 1/k$ (dihedral) symmetry of the correlator, and one observes that the non-vanishing coefficients can be expressed in terms of $\alpha$. The first few terms read
\begin{multline}\label{Fzetaexp}
F(k) = \log{\left[\frac{\pi ^2}{4}\right]}
+ [2 (3-4 \F) \F-\frac{1}{2} \log (32 \F) ] \zeta^2
+ \frac{\F}{6} [ 3-2 \F (9+4 \F (-5+6 \F)) ] \zeta ^4 +\mathcal{O}(\zeta^6)\, ,
\end{multline}
which is thus an even regular series, up to the tails of logarithms encoded in eq.~\eqref{eq:alpha}.  We remark that the expansion~\eqref{Fzetaexp} has a very small range of validity if $\F$ is given by eq.~\eqref{eq:alpha}.

\subsection{Comparison with periodic fishnets}\label{sec:continuum}

It is instructive to compare our formula with the large-order behaviour of fishnet graphs with doubly periodic boundary conditions. In his pioneering paper, A.~Zamolodchikov \cite{Zamolodchikov:1980mb} obtained the expression for the $m\times n$ fishnet torus partition function $Z_{m, n}$, in any spacetime dimensions $D$, using integrability and bootstrap conditions. Assuming the thermodynamic scaling, he found that
\beq\label{eq:Z-scaling}
\lim_{m, n\rightarrow \infty} \frac{1}{mn}\log{Z_{m, n}} = -\log{g_c^2}\, ,
\eeq
where the so-called critical coupling $g_c$ is a constant which depends on $D$ but not on the fishnet lengths $m, n$. The explicit expression for $D=4$ is given by \cite{Zamolodchikov:1980mb}
\beq
-\log{g_c^2} = \log{\left[\frac{\Gamma(\tfrac{1}{4})^4}{32\pi}\right]} = 0.541...\, .
\eeq
Clearly, this result bears no similarity at all with our general expression for $F(k)$. It is also numerically different, over the whole physical range $k\in (1, \infty)$; it is just over half of the lowest value achieved by $F(k)$, namely $F(1) = \log{(\pi^2/4)} = 0.903...$.

From the 2d lattice perspective, this discrepancy and, more generally, the dependence of the scaling function $F$ on the aspect ratio $k$ appear surprising at first sight. They go against the common wisdom that the thermodynamic free energy is extensive and independent of the boundary conditions.%
\footnote{They are also at odds with common beliefs about universality of large-order behaviour of Feynman diagrams~\cite{Parisi:1977kw,Parisi:1976zh}.}
Put differently, they suggest that the boundary conditions set by the four-point fishnet correlator are atypical and ``severe'' enough to penetrate deeply into the bulk of the lattice, in such a way that the uniform periodic distribution may hold only locally, far away from the boundary, if at all.

This sensitivity to the boundary conditions is likely to hold for other correlators. In fact a similar phenomenon may be observed for half-periodic fishnets, mixing open and closed boundary conditions, such as the one shown in the right panel of figure~\ref{BCs}. An interesting aspect of these mixed boundary conditions is that they provide a regularization of the otherwise ill-defined doubly periodic fishnets, which suffer from UV/IR divergences. They yield finite correlation functions, which depend on several cross ratios and which may be expanded over a complete basis of states in principal series representations, as discussed in detail in ref.~\cite{Grabner:2017pgm} in a particular case. Doing so, one projects over states carrying arbitrary scaling dimension $\Delta$ along the radial direction.%
\footnote{Strictly speaking, $\Delta = 2+\I \nu$ with $\nu \in \mathbb{R}$ for principal series representations and other values of $\Delta$ are reached by analytic continuation.}. Fishnets of this type were considered in the thermodynamic limit in ref.~\cite{Basso:2018agi}, for certain states, and their free energy was calculated for a large range of scaling dimensions $\Delta$ using the Thermodynamic Bethe Ansatz (TBA) equations of the fishnet theory. Importantly, they were found to scale as in~\eqref{eq:Z-scaling} only for $\Delta$'s that are much smaller than the fishnet size. The latter requirement is understood as a condition for the validity of the continuum description of large fishnet graphs, which is governed in the case at hand by the 2d non-linear $AdS_{5}$ sigma model. If on the contrary the dimension $\Delta$ scales large with the system size, then the corresponding state in the sigma model is very excited, with a finite energy density, and the free energy density departs significantly from the ground-state behavior~\eqref{eq:Z-scaling}.

Another way of phrasing the situation is in terms of the so-called graph building operator $T_{m}$~\cite{Gromov:2017cja,Zamolodchikov:1980mb}, which is the integral kernel which adds a rung to periodic fishnet graphs with $m$ radial lines. It may be interpreted as a finite-time evolution operator for an integrable spin chain, with spins in a suitable representation of the four-dimensional conformal group~\cite{Gromov:2017cja,Derkachov:2020zvv}. Its logarithm defines a Hamiltonian for a system of length $m$, which is well approximated by the Hamiltonian of the $AdS_5$ sigma model,
\beq\label{eq:low-energy}
-\log{T_{m}} = m\log{g^2_{c}} + H_{AdS} + \ldots\, ,
\eeq 
in the large volume limit $m\rightarrow \infty$, up to higher-derivative interactions associated with $1/m$ corrections. The point is that this expansion can be trusted, and the latter corrections neglected, only when the quantum numbers of the state, and in particular the scaling dimension, are much smaller than the length $m$. One may expect this condition to be observed for $m\times n$ half-periodic fishnets, at generic values of the cross ratios, as long as both $m$ and $n$ are large -- with the limit of large number of rungs (or large discrete time), $n \rightarrow \infty$, being used here to project on the 2d low-energy states. However, it is not obeyed, even for large $m$ and large $n$, for the special configurations of the boundary points which do not have a smooth 2d limit and extract contributions from the high-energy part of the spectrum of $\log{T_m}$. 

With that in mind, it appears less surprising to find that the boundary conditions set by our four-point function are strong enough to drive the system away from the scaling~\eqref{eq:Z-scaling}. Indeed, the fishnet lengths $m, n$ coincide in this case with the scaling dimensions of the operators sitting at the boundary, which therefore scale large in the thermodynamic limit. In other words, the interpretation would be that it is because of the ``roughness'' of the boundary conditions that we observe a loss of universality in the thermodynamic scaling.

We should stress that strong sensitivity to the boundary conditions has also been observed in other solvable statistical models. The most famous example is given by the six-vertex model and related exactly solvable models of statistical mechanics, see~\cite{Korepin:2000,Zinn-Justin:2002} and references therein, where the partition function with domain-wall boundary conditions was found to scale very differently from the torus partition function in the large volume limit. The interpretation here is that the former partition function is subject to the formation of a so-called arctic curve~\cite{Jockusch,Cohn}, which separates two different phases of the model. Namely, the configurations appear nearly frozen across a macroscopic domain near the boundary and only start fluctuating deep inside the lattice, where they approach the disordered phase. The free energy is not distributed evenly throughout the lattice but instead undergoes a (more or less) sharp transition at the place where the two phases co-exist. It has also been suggested that this phenomenon could result from the roughness of the boundary conditions~\cite{Korepin:2000,Zinn-Justin:2002} at least in some situations.

It would be interesting to see if this analogy can be made more precise for the fishnet lattices. It would also be interesting to see if the phenomenon discussed here admits a dual AdS interpretation, despite the tension with the 2d low-energy description~\eqref{eq:low-energy}. In the next section, we will see some indirect evidence of a connection with classical string theory in $AdS_{3}$, as we consider a generalized thermodynamic scaling limit which combines large fishnet and short spacetime limits.

\section{Short-distance limit and spinning string}\label{sec:spinning}

Let us now discuss a more general scaling limit, in which the spacetime parameters scale large with the fishnet lengths $m, n$. We shall focus on the situation where $|\sigmam| \sim m$, which corresponds to the Euclidean short-distance regime, $x_{1}\rightarrow x_{3}$ or $x_{2}\rightarrow x_{3}$, depending on the sign of $\sigma$, and use the dual integral~\eqref{Jmn}.%
\footnote{We could also consider a rescaling of $\sigmap$, along with $\sigmam$. We do not expect any difference as long as $|\sigmap| < |\sigmam|$. Indeed, increasing |$\varphi|$ merely shifts the linear part of the potential, which becomes piecewise linear, with a plateau up to $x \sim |\sigmap|/m$ followed by the linear potential. This plateau lies outside of the saddle-point distribution for $|\sigmap|< |\sigmam|$, thus leaving the analysis unchanged. The alternative situation where $|\sigmap| > |\sigma|$ is more delicate and relates to the lightlike short-distance limit of the correlator.}

\subsection{Mapping with spinning string}

The dual equations stay essentially the same in the general scaling limit, if not for the potential~\eqref{eq:scaledV}, which is replaced by
\beq
V(x) \approx - \frac{1}{\beta} \log{(x^2-\sigma^2)} + \frac{|x|}{m}\, .
\eeq
Introducing the density~\eqref{constraint} and the parameter
\beq
\xi = \frac{\beta |\sigma|}{4\pi m}\, ,
\eeq
which is held fixed, with $\beta = 1/(k-1)$, in the limit $|\sigma|, m, n \rightarrow \infty$, the saddle-point equation~\eqref{saddle-point} becomes
\beq\label{eq:string-eq}
0 = \frac{x}{x^2-\xi^2} -2\pi + \dashint_{a}^{b} \frac{4x \rho(y)\rmd y}{x^2-y^2}\, , \qquad x\in (a, b)\, ,
\eeq
with $\xi \leqslant a \leqslant b$. Our previous discussion relates then to the limit $\xi \rightarrow 0$. 

Now, equation~\eqref{eq:string-eq} turns out to be the same as the finite-gap equation \cite{Kazakov:2004nh,Casteill:2007ct,Belitsky:2006en} for a classical string in $AdS_{3}\times S^{1}$, constructed explicitly by Frolov and Tseytlin \cite{Frolov:2002av}.%
\footnote{Equation~\eqref{eq:string-eq} also shares similarities with the equation for a model of open strings in $D=0$ dimensions studied in ref.~\cite{Kazakov:1989cq}.} It describes a folded string moving in $AdS_{3}$ with spin $S$ and global time energy $E$ and boosted with a momentum $J$ along a big circle in $S^1$. In this setup the parameter $\xi$ plays the role of the so-called BMN coupling,
\beq\label{eq:lambda}
\xi^2 = \lambda/(4\pi J)^2\, ,
\eeq
with $\sqrt{\lambda}/2\pi$ the string tension and with $\lambda$ the 't Hooft coupling of the dual gauge theory. At weak BMN coupling, the string theory description merges smoothly with the spin-chain analysis reviewed in section \ref{sec:spin-chain}, as discussed in detail in refs.~\cite{Kazakov:2004nh,Beisert:2003ea, Beisert:2003xu}. However, as the coupling gets bigger, the classical string analysis takes over.

The solution to eq.~\eqref{eq:string-eq} is known for any $\xi$ and may be found explicitly in refs.~\cite{Casteill:2007ct,Belitsky:2006en,Kazakov:2004nh}. Here, we simply quote the expressions for the parameters of the distribution, 
\beq\label{eq:string-parameters}
b = 4K(q) \sqrt{(a^2-\xi^2)(b^2-\xi^2)}\, , \qquad \beta = 2b E(q)-\frac{1}{2}-\frac{2\xi^2 K(q)}{b}\, ,
\eeq
and $q=1-a^2/b^2 \in (0, 1)$, which generalize to $\xi \neq 0$ the formulae in eqs.~\eqref{eqn-DiffEqnEllipPara}. Recall that $\beta$ controls the normalization of the density in our convention. In the string theory, it relates to a linear combination of the string quantum numbers~\cite{Casteill:2007ct,Belitsky:2006en,Kazakov:2004nh}
\beq\label{eq:string-beta}
\beta = 2\int\limits_{a}^{b}\rmd x \rho(x) = \frac{E+S-J}{2J}\, .
\eeq
It reduces to the spin-chain expression, see comment below~\eqref{eq:rho-spin-chain}, in the weak BMN coupling limit $\xi \rightarrow 0$, that is, when the ``anomalous dimension'' $E-S-J \sim 0$.

\subsection{Short-string regime}

One may then derive a differential equation for the scaling function,
\beq
f = f(\beta, \xi) = \lim_{m, n, |\sigmam|\rightarrow \infty} m^{-2} \log{\Phi_{m, n}}\, ,
\eeq
by following the same lines of analysis as before. One only needs to replace the potential used in eq.~\eqref{eq:firstly} by its deformed version,
\beq
4\int\limits_{a}^{b}\rmd x \delta\rho(x) (\log{\sqrt{x^2-\xi^2}}-2\pi  x) = 2\log{\left[\frac{\sqrt{a^2-\xi^2}+\sqrt{b^2-\xi^2}}{2}\right]}-8bE(q)\, ,
\eeq
with $\delta \rho = \partial_{\beta}\rho$ given in eq.~\eqref{eq:exp-delta-rho}. It yields the differential equation
\beq\label{eq:string-diff}
\frac{\partial}{\partial\beta} (\beta^2(f-\Cp-4\pi \xi/\beta)) = 2\beta\log{\left[\frac{b^2-a^2}{4}\right]} + 2\log{\left[\frac{\sqrt{a^2-\xi^2}+\sqrt{b^2-\xi^2}}{2}\right]}-8bE(q)\, ,
\eeq
with the derivative taken at fixed $\xi$ and with $\Cp$ given explicitly in eq.~\eqref{Cprime}. The extra piece on the left-hand side stems from the prefactor of the dual integral~\eqref{eqn-detInt},
\beq
d(z, \bar{z})^m \approx e^{m |\sigma|} = e^{4\pi m^2 \xi/\beta}\, ,
\eeq
in the limit $m, |\sigma|\rightarrow \infty$.

Integrating eq.~\eqref{eq:string-diff} for general $\beta$ and $\xi$ is beyond the scope of this paper. Here, we will focus on taking the strong coupling limit $\xi \gg 1$ at fixed $\beta$. This limit maps to the short-string domain on the string theory side, with the spin, energy and momentum obeying the flat-space dispersion relation $S\sim E^2-J^2$~\cite{Frolov:2002av}.%
\footnote{The scaling is easily understood from eqs.~\eqref{eq:lambda} and~\eqref{eq:string-beta}. At large $\xi$, the sphere momentum $J$ is small (in string units). Hence, in order to keep $\beta = \mathcal{O}(1)$, one must have $E \sim J$, meaning that the energy is small as well.}
The small and large $\beta$ limit coincides then with $J^2 \gg S$ and $J^2 \ll S$, respectively.

At the level of eq.~\eqref{eq:string-eq}, the short-string regime corresponds to an expansion around the bottom of the potential $V'(c) = 0$ in the limit $\xi \rightarrow \infty$. Expanding eqs.~\eqref{eq:string-parameters} around this point yields
\beq\label{eq:abc}
a = c + \frac{\beta - \sqrt{\beta (1+\beta)}}{2\pi} + \mathcal{O}(1/\xi) \, , \qquad b = c + \frac{\beta + \sqrt{\beta (1+\beta)}}{2\pi} + \mathcal{O}(1/\xi)\, ,
\eeq
where $c = (1+\sqrt{1+16\pi^2 \xi^2})/4\pi \sim  \xi$. Inserting the expressions in the right-hand side of the differential equation~\eqref{eq:string-diff}, using
\beq
\begin{aligned}
&2\beta\log{\left[\frac{b^2-a^2}{4}\right]} = \beta \log{\left[\frac{\beta (1+\beta)\xi^2}{4\pi^2}\right]}+ \mathcal{O}(1/\xi)\, , \\
&2\log{\left[\frac{\sqrt{a^2-\xi^2}+\sqrt{b^2-\xi^2}}{2}\right]}-8bE(q) = -4\pi \xi + \log{\left[\frac{(1+\beta)\xi}{2\pi}\right]} -2\beta -1 + \mathcal{O}(1/\xi)\, ,
\end{aligned}
\eeq
one finds that the terms linear in $\xi$ cancel out, such that
\beq
\frac{\partial}{\partial\beta} (\beta^2(f-\Cp))
= (1+2\beta) \log[\xi/(2\pi e)] + (1+\beta) \log{(1+\beta)}  + \beta \log{\beta} + \mathcal{O}(1/\xi)\, .
\eeq
This equation is significantly simpler than the one encountered earlier, for small $\xi$, and it can be integrated directly,
\beq\label{eq:stringy-f}
f  = k\log{(8\pi (k-1)\xi)}
+ \frac{1}{2} \Big[ k^2\log{k}-(k+1)^2\log{(k+1)} \Big] + \frac{3k}{2} + \mathcal{O}(1/\xi)\, ,
\eeq
using eq.~\eqref{Cprime} for $\Cp$ and the boundary condition $\lim_{\beta\rightarrow 0}(\beta^2 f) = 0$.
Re-expressing it in terms of the original variables, we find
\beq\label{eq:stringy}
\log{\Phi_{m, n}} = mn\log{(2|\sigma|)}+\frac{1}{2}(m^2\log{m} + n^2 \log{n}-(m+n)^2 \log{(m+n)})+ \frac{3 m n}{2} + \mathcal{O}(1/|\sigma|)\, ,
\eeq
which holds when $|\sigma| \gg m, n \gg 1$. It is manifestly symmetric under $m\leftrightarrow n$. One reads from it that $\Phi_{m, n} \sim |\sigma|^{mn}$ in the short-distance limit, $|\sigma| \rightarrow \infty$. Note that it sends the exponent infinitely far away from the predicted one~\eqref{eq:Z-scaling} for periodic fishnets. This scaling is nonetheless in perfect agreement with the UV power counting of the fishnet integral, with each loop giving rise to a logarithmic divergence $\sim |\sigma|$. It is quite intriguing to see this simple field theory behaviour arising here from a short string analysis. 

Let us add finally that subleading terms can be easily produced by keeping higher orders in~\eqref{eq:abc}. For illustration, one easily finds
\beq\label{eq:delta-stringy}
\delta \log{\Phi_{m, n}} = \frac{mn(m+n)}{2 |\sigma|} - \frac{(m n)^2}{4\sigma^2} - \frac{mn(m+n)(m^2+mn+n^2)}{24|\sigma|^3}  + \mathcal{O}(1/\sigma^4)\, ,
\eeq
for the first few corrections to~\eqref{eq:stringy}. They constitute corrections to the flat-space regime, coming from the curvature of AdS, in the string theory interpretation.

\subsection{Hankel determinant of factorials}

We can compare the ``stringy'' prediction~\eqref{eq:stringy} with a direct evaluation of the determinant of ladders. As we will now explain, the latter simplifies drastically at large $\sigma$ and can be given in closed form for all $m$ and $n$. We first note the large $\sigma$ behaviour of the ladders, choosing $\sigma \rightarrow -\infty$ for convenience, 
\beq
L_{p}(z, \bar{z}) \approx \frac{(2|\sigma|)^p}{p!} (z-\bar{z})\, .
\eeq
The highest power of $|\sigma| = \tfrac{1}{2}[-\log{(z\bar{z})}]$ dominates in the sum~\eqref{ladders}, using the fact that $\textrm{Li}_{j}(z) \approx z$ in the limit $|\sigma| = -\sigma \rightarrow \infty$, or $z, \bar{z} \ll 1$. Inserting the above estimate inside the determinant~\eqref{eq:I-det} then yields the simple expression
\beq\label{eq:Hankel-fact}
\Phi_{m, n} \approx \frac{(2|\sigma|)^{m n}}{\mathcal{N}} \det ((i+j+n-m-2)!)_{1\leqslant i, j \leqslant m}\, ,
\eeq
for the leading large $|\sigma|$ behaviour.

Now, determinants of Hankel matrices of factorials form nice sequences, which can be written concisely in terms of Barnes' $G$-function, see eq.~\eqref{eq:G-barnes}. In particular, a classic mathematical result yields for $n=m$,
\beq
\det\, ((i+j-2)!)_{1\leqslant i, j \leqslant m} = G(m+1)^2 = \prod_{k=0}^{m-1} (k!)^2\, .
\eeq
The general formula, valid for any $n\geqslant m$, is given by \cite{BASOR2001214}
\beq
\det ((i+j+n-m-2)!)_{1\leqslant i, j \leqslant m} = \frac{G(m+1)G(n+1)}{G(n-m+1)}\,,
\eeq
and it can be derived by evaluating the determinant with the method of orthogonal polynomials, as explained in appendix~\ref{sec:orthogonal}. It combines neatly with the prefactor~\eqref{eq:G-barnes}
such as to give
\beq\label{eq:Phi-0}
\Phi_{m, n} \approx (2|\sigma|)^{m n} \frac{G(m+1)G(n+1)}{G(n+m+1)}\, ,
\eeq
which is symmetric under $m\leftrightarrow n$, as it should be. The large $m, n$ limit follows immediately from the asymptotic behaviour of Barnes' $G$-function at large argument, eq.~\eqref{eq:G-large}, reproducing perfectly the stringy result~\eqref{eq:stringy}.

One may also find concise expressions for the subleading logarithmic corrections at large $|\sigma|$. They originate from the lower terms in the polynomials in $|\sigma|$ in the ladders,
\beq
L_{p} \rightarrow (z-\bar{z}) \sum_{j=p}^{2p}\frac{j!(2|\sigma|)^{2p-j}}{p!(j-p)! (2p-j)!} =  \frac{(z-\bar{z})}{\sqrt{\pi} p!} e^{|\sigma|} (2|\sigma|)^{\frac{1}{2}+p} K_{\frac{1}{2}+p}(|\sigma|)\, ,
\eeq
with $K$ the modified Bessel function of the second kind. Plugging this expression inside the determinant and expanding at large $\sigma$, for several values of $m, n$, we found that the corrections can be parametrized in terms of symmetric polynomials in $m$ and $n$ of increasing degrees,
\beq
\log{(\Phi_{m, n}/\Phi^{0}_{m, n})} = \frac{mn(m+n)}{2|\sigma|}-\frac{mn (mn+1)}{4\sigma^2} - \frac{m n (m+n) (m^2+mn +n^2-7)}{24|\sigma|^3} +\ldots \, ,
\eeq
with $\Phi_{m, n}^{0}$ denoting the leading order expression~\eqref{eq:Phi-0}. These polynomials are readily seen to match with the stringy curvature corrections~\eqref{eq:delta-stringy}, when $m, n \gg 1$.

\section{Conclusion}\label{sec:conclusion}

In this paper, we examined simple four-point fishnet integrals in four dimensions, for operators carrying finite or large scaling dimensions, $m, n$. The equivalence between the various integral representations allowed us to obtain closed-form expressions for the free-energy density controlling the large-order behaviour of these diagrams. It was found to be independent of the cross ratios (away from singular points), in line with common wisdom on the scaling of boundary data in the thermodynamic limit. However, the ``universality property'' of the free energy appears incomplete, owing to non-trivial dependence on the fishnet aspect ratio $k = n/m$.

The dependence on $k$ of the free energy raises a number of questions. In particular, it suggests that the fishnet partition function studied in this paper may be subject to the formation of an arctic curve, as seen in the six-vertex model and related exactly solvable models of statistical mechanics~\cite{Jockusch,Cohn,Korepin:2000,Zinn-Justin:2002} with domain wall boundary conditions~\cite{Korepin:1993kvr}. The analogy is quite neat at the mathematical level, with the latter set of boundary conditions giving rise to a determinant, while the periodic ones are found to be governed by a more involved (linearized) TBA equation.

It would be interesting to see if this analogy holds up under a more detailed analysis. Our measure of the free energy density is rather crude, since it averages over the entire graph. It would be interesting to dive more deeply into the diagram and extract information about the local distribution. This may perhaps be done by deforming the boundary conditions, adding excitations on the local operators at the boundary, or by considering higher-point functions, with additional operators. (For instance, a five-point function with a Lagrangian insertion might help exploring local properties of the fishnet graphs.) One may also draw inspiration from the methods used to construct the arctic curve in the six-vertex model, see e.g.~refs.~\cite{Colomo:2007,Colomo:2008zz}.

Recently, a lot of progress has been made in obtaining alternative descriptions of the fishnet diagrams in terms of dual systems in AdS. In particular, we mentioned that cylindrical fishnet graphs are expected to relate to the non-linear AdS sigma model in the continuum limit. In light of the aforementioned discrepancy, it is not immediately clear if this description applies to the thermodynamic limit of the open-string-like correlators studied in this paper. The question remains as to whether a dual AdS description exists in these cases as well. This description may not be smooth throughout the entire system or given in terms of a local 2d quantum field theory, but it may nonetheless be formulated entirely in the AdS space.
The AdS string-bit formulation of the fishnet graphs developed recently in refs.~\cite{Gromov:2019aku,Gromov:2019bsj,Gromov:2019jfh,Gromov:2021ahm} -- dubbed the fish-chain model -- may shed light on it, as it is does not assume a low-energy approximation and is naturally tied to the regime of large scaling dimensions.
Still, some work appears needed to cast it in a form that is suitable for studying the thermodynamic limit.

The generalized scaling limit discussed in section \ref{sec:spinning} is an interesting corner for exploring a potential new connection with a dual AdS description. In this case we saw that the saddle-point equation agrees perfectly with the finite-gap equation for a rigid string spinning in $AdS_{1}\times S^{1}$. Some aspects of the mapping, such as the emergence of the internal circle $S^{1}$ or the relation between the spacetime parameter of the fishnet correlator and the BMN coupling of the string, are quite intriguing and it is not yet clear if this mapping results from a mathematical accident or stands as an indication of a genuine correspondence.

\acknowledgments

We thank Paul Fendley, Volodya Kazakov, Shota Komatsu, Ivan Kostov, Jorge Kurchan, Oliver Schnetz, Didina Serban, Amit Sever and F\'elix Werner for interesting discussions, and Volodya Kazakov and Ivan Kostov for useful comments on the manuscript. The research of BB, DK and DlZ was supported by the French National Agency for Research, under the grants ANR-17-CE31-0001-01 and ANR-17-CE31-0001-02. The research of LD was supported by the US Department of Energy under contract DE--AC02--76SF00515. AK acknowledges support from the ANR grant ANR-17-CE30-0027-01 RaMaTraF and from ERC under Consolidator grant number 771536 (NEMO). The work of DlZ was supported in part by a center of excellence funded by the Israel Science Foundation (grant number 2289/18) and by the Israel Science Foundation (grant number 1197/20).
LD and DK thank the Kavli Institute for Theoretical Physics for hospitality at the beginning of this project (National Science Foundation Grant No. NSF PHY-1748958). DlZ is grateful to CERN for the warm hospitality in the last stage of this project.

\appendix

\section{Technical lemmas}\label{sec:lemmas}

In this appendix, we provide the key lemmas used for the exact evaluation of the BMN and FT integrals.

The following Lemma yields an integral relation for the product of two determinants defined by two sets of functions. 
    \begin{lemma}[Cauchy-Binet-Andr\'eief formula]
\label{lemma:andreief}
For a set of functions $\lbrace f_i, g_j \rbrace_{i,j\in [1,n]}$ and measure $\rmd \nu$, we have the Cauchy-Binet-Andr\'eief formula
\begin{equation}
\int_{\mathbb{R}^n}\frac{\rmd \nu( \bm{v})}{n!} \det (f_i(v_j))\det(g_i(v_j))=\det\left[\int_\mathbb{R}\rmd \nu(v)f_i(v)g_j(v)\right]\, .
\end{equation}
\end{lemma}
~
\begin{proof}
See refs.~\cite{andreief1883note,forrester2018meet}.
\end{proof}~\\

The next Lemma permits one to convert sums over discrete integers into contour integrals.
\begin{lemma}[Mellin-Barnes summation formula]\label{lemma:MBclassique}
Let $\mathcal{C}=\varepsilon+i\mathbb{R}$ for some $\varepsilon \in(0,1)$ be a contour in the complex plane, parallel to the imaginary axis and $y\in \mathbb{C}$ such that $\textrm{Re} \, y>0$. The Mellin-Barnes summation formula allows one to replace a summation over integers by an integral over $\mathcal{C}$,
\begin{equation}\label{eqn:MB}
\sum_{a\geqslant 1} (-y)^a f(a)=-\int_{\mathbb{R}}\mathrm{d}\r\, \frac{y}{y+e^{-\r}}\int_{\mathcal{C}}\frac{\mathrm{d}w}{2\pi \I}e^{-w\r}f(w)\, .
\end{equation}
\end{lemma}
For completeness, we provide a short non-rigorous proof of the classical Mellin-Barnes summation formula, see ref.~\cite[Lemma 3.2.13]{borodin2014macdonald}  for convergence issues and conditions on the function $f$.

~
\begin{proof}
By a residue calculus (and ignoring convergence issue), we note the identity
\begin{equation}
\sum_{a\geqslant 1} (-y)^a f(a)=-\int_{\mathcal{C}}\frac{\mathrm{d}w}{2\pi \I}\frac{\pi}{\sin{(\pi w)}} y^w f(w)\, ,
\end{equation}
using ${\rm Res}_{w=a}[ \pi / \sin{(\pi w)}]=(-1)^{a}$. Furthermore, we note the second identity for $\textrm{Re} \, y>0$ and $w \in \mathcal{C}$,
\begin{equation}\label{eq:int-sine}
\frac{\pi }{\sin{(\pi w)}}y^w=\int_\mathbb{R} \rmd \r \frac{y}{y+e^{-\r}}e^{-w\r}\, ,
\end{equation}
where the right-hand side can be viewed as a Mellin transform of a \textit{Fermi} factor. 
\end{proof}~\\

We introduce now a lemma about derivatives of the inverse of hyperbolic cosines which will be essential to prove Lemma \ref{lemma:vandermonde}.
\begin{lemma}[Derivative of inverse of hyperbolic cosine]
\label{lemma_polynomial}
For all $j\geqslant 0$, the $j$-th derivative of the inverse of hyperbolic cosine reads
\begin{equation}
    \frac{\partial^{j-1}}{\partial u^{j-1}} \sech{(\pi u)}=\sech{(\pi u)} P_{j-1}(\tanh{(\pi u)})\, ,
\end{equation}
where $P_k$ is a polynomial of leading order $k$ and leading coefficient $a_k=(-\pi)^k k!$.
\end{lemma}
~
\begin{proof}
 We prove it by induction. Assume it is valid for some $j\geqslant 1$, we get

\begin{equation}
    \frac{\partial}{\partial u}\frac{P_{j-1}(\tanh{(\pi u)})}{\cosh{(\pi u)}}=\pi\frac{ \tanh'{(\pi u)}P_{j-1}'(\tanh{(\pi u)})-P_{j-1}(\tanh{(\pi u)})\tanh{(\pi u)}}{\cosh{(\pi u)}}\, .
\end{equation}
    Using the expression of the derivative of the hyperbolic tangent, $\tanh'=1-\tanh^2$, we define $P_j$ as
    \begin{equation}
        P_j(x)=\pi[ (1-x^2)P_{j-1}'(x)- xP_{j-1}(x)]\, .
    \end{equation}
The leading coefficient of $P_j$ is then obtained as $a_j=-\pi j a_{j-1}$.    Hence, by induction, as $a_0=1$, for all $j\geqslant 0$, $a_j=(-\pi)^j j!$.
\end{proof}~\\

The next Lemma allows us to transform a Vandermonde determinant over hyperbolic tangents into a Vandermonde determinant over partial derivatives. 
\begin{lemma}[Equivalent Vandermonde representations]
For any set of variables $\lbrace u_j \rbrace_{j\in [1,n]}$, the following equality holds
\label{lemma:vandermonde}
\begin{equation}
  (-\pi)^{\frac{1}{2}n(n-1)} \prod_{j=1}^n (j-1)! \prod_{i=1}^n \sech{(\pi u_i)}\prod_{i<j}(\tanh{(\pi u_j)}-\tanh{(\pi u_i)})=\Delta_{n}(\partial_{u})\prod_{i=1}^n \sech{(\pi u_i)}\, ,
\end{equation}
where $\Delta_{n}(\partial_{u}) = \prod_{i<j}(\partial_{u_j}-\partial_{u_i})$.
\end{lemma}
~
\begin{proof}
Starting from the partial-derivative Vandermonde representation and Lemma \ref{lemma_polynomial}, we have
\begin{equation}
    \Delta_{n}(\partial_{u})\prod_{i=1}^n \sech{(\pi u_i)}=\det_n \bigg[\frac{\partial^{j-1}}{\partial u_i^{j-1}} \sech{(\pi u_i)}\bigg]=\det_n \bigg[ \frac{P_{j-1}(\tanh{(\pi u_i)})}{\cosh{(\pi u_i)}}\bigg]\, .
\end{equation}
Combining the rows of the matrix, we can retain only the highest order term of each polynomial $P_j$, 
    \begin{equation}
    \Delta_{n}(\partial_{u})\prod_{i=1}^n \sech{(\pi u_i)}=\prod_{j=1}^n a_{j-1} \times \det_n \bigg[ \frac{\tanh{(\pi u_i)}^{j-1}}{\cosh{(\pi u_i)}}\bigg]\, .
    \end{equation}
Lastly, we observe that the remaining determinant is again a Vandermonde determinant with hyperbolic tangents as arguments,
    \begin{equation}
         \Delta_{n}(\partial_{u})\prod_{i=1}^n \sech{(\pi u_i)}=(-\pi)^{\frac{1}{2}n(n-1)}\prod_{j=1}^n (j-1)!\prod_{i=1}^n \sech{(\pi u_i)}\prod_{i<j}(\tanh{(\pi u_j)}-\tanh{(\pi u_i)})\, .
    \end{equation}
    \end{proof}

\section{Elliptic parameters} \label{sec:parameter}

In this appendix, we determine the parameters $A,w_2,w_3,w_4$ in the elliptic parametrization \eqref{eqn-Ellip-z}, \eqref{eqn-Ellip-gz} by fixing the mapping at the boundary points.

\textit{Matching the asymptotics.} The asymptotic behaviours at small $u$ and at large $u$ are easily matched on both sides of eq.~\eqref{eqn-Ellip-gz}. For $u$ small, $w$ is small and in this limit  eq.~\eqref{eqn-Ellip-gz} yields
\be
u = (1+k) \sqrt{w} + \mathcal{O}(w) \, ,
\ee
while eq.~\eqref{eqn-Ellip-z} gives
\be
u = - \frac{2 A w_3}{\sqrt{w_2 w_4}} \sqrt{w} + \mathcal{O}(w) \, .
\ee
Hence,
\be
1+k = - \frac{2 A w_3}{\sqrt{w_2 w_4}} \, .
\ee
We can proceed similarly for the large $u$ regime, which maps to large $w$. Equation~\eqref{eqn-Ellip-gz} yields
\be
\frac{1}{\sqrt{w}} = \frac{k-1}{u}  + \mathcal{O}(1/u^3) \qquad \Rightarrow \qquad u = (k-1) \sqrt{w} +  \mathcal{O}(w^{-1/2}) \, ,
\ee
while eq.~\eqref{eqn-Ellip-z} gives
\be
u = 2 A \sqrt{w} +  \mathcal{O}(w^{-1/2}) \, .
\ee
Therefore,
\be
A = \frac{k-1}{2} \, .
\ee

\textit{Matching boundary points.} The remaining parameters are fixed by evaluating the solution at special points on the boundary. Namely, enforcing that the pre-images of $\{0, w_2, w_3, w_4\}$ are $\{0,\frac{\I}{2}-\I0,\frac{\I}{2}+B,\frac{\I}{2}+\I 0\}$, we obtain the relations
\bea
\frac{1}{2} & = A \int_{w_2}^0 \rmd t \frac{t - w_3}{\sqrt{(-t)(t-w_2)(t-w_4)}} \, , \\
B & = A \int_{w_3}^{w_2} \rmd t \frac{t - w_3}{\sqrt{(-t)(w_2-t)(t-w_4)}} \, , \\
B & = A \int_{w_4}^{w_3} \rmd t \frac{w_3 - t}{\sqrt{(-t)(w_2-t)(t-w_4)}} \, .
\eea
Taking the difference of the last two equations, we get
\be \label{eqn-Ellip-1}
w_4 E(q) = w_3 K(q), \qquad q \equiv 1 - \frac{w_2}{w_4} \in [0,1) \, ,
\ee
while the first equation yields
\be \label{eqn-Ellip-2}
\frac{1}{4 A \sqrt{-w_4}} = E(p) + \Big( \frac{w_3}{w_4}-1 \Big) K(p), \qquad p \equiv \frac{w_2}{w_4} \, ,
\ee
with $K$ and $E$ the complete elliptic integrals of the 1st and 2nd kind, see eq.~\eqref{eqn-defEK}.

Equation \eqref{eqn-Ellip-2} can be further simplified by eliminating $w_3/w_4$ with the help of eq.~\eqref{eqn-Ellip-1}. It gives
\bea
\frac{1}{4 A \sqrt{-w_4}} = \frac{E(p)K(q)+K(p)E(q)-K(p)K(q)}{K(q)}  = \frac{\pi}{2 K(q)}\, ,
\eea
using the Legendre relation for elliptic functions of conjugate moduli, $p+q=1$, to simplify the numerator.

\textit{Summary.} We found
\bea 
A & = \frac{k-1}{2} = \frac{\sqrt{1-q} K(q)}{E(q)-\sqrt{1-q} K(q)}, \\
-w_2 & = (1-q)(-w_4) =\frac{1}{4\pi^2} [E(q)-\sqrt{1-q} K(q)]^2, \\
-w_3 & = \frac{K(q)E(q)}{\pi^2 (k-1)^2} = \frac{1}{4\pi^2} \frac{E(q)}{K(q)} \frac{[E(q)-\sqrt{1-q} K(q)]^2}{1-q}, \\
-w_4 & = \frac{K^2(q)}{\pi^2 (k-1)^2} = \frac{1}{4\pi^2} \frac{[E(q)-\sqrt{1-q} K(q)]^2}{1-q}, \\
B & = (k-1)\bigg[ \frac{w_3}{\sqrt{-w_4}} F(\gamma,q) + \sqrt{-w_4} E(\gamma,q) \bigg], \qquad \gamma = \arcsin \sqrt{\frac{w_3 - w_4}{w_2 - w_4}}, \\
\eea
where $E(\gamma,q), F(\gamma,q)$ are the incomplete elliptic integrals obtained by changing the upper limit in eq.~\eqref{eqn-defEK} from $\pi/2$ to $\gamma$. We arrive at eq.~\eqref{eqn-Ellip-Final} by replacing the elliptic integrals by the parameters $a,b, \beta$ introduced in eq.~\eqref{eqn-DiffEqnEllipPara}.

\section{Orthogonal polynomials} \label{sec:orthogonal}

In this appendix, we calculate the short-distance limit of the determinant of ladders, eq.~\eqref{eq:Phi-0}, using the method of orthogonal polynomials, see ref.~\cite{DiFrancesco:1993cyw} for a review. First, recall the integral representation~\eqref{eq:ladder-intermediate} of the ladder $L_{p}(z, \bar{z})$. In the limit $\sigmam \rightarrow -\infty$, or equivalently $z, \bar{z} \rightarrow 0$, it yields
\beq
M_{p} = p! (p-1)! L_{p} \approx (z-\bar{z}) \times (2|\sigma|)^{p} \int_{0}^{\infty} \rmd r\, r^{p-1} e^{-r}\, ,
\eeq
such that the correlator~\eqref{eq:Phi-I} can be written as
\beq
\Phi_{m, n}  \approx \frac{(2|\sigma|)^{mn}}{\C} \det \left[ \int_{0}^{\infty} \rmd r\, r^{\ell+i+j-2} e^{-r} \right]_{1\leqslant i, j \leqslant m}\, ,
\eeq
with $\ell\equiv n-m$. One then introduces the basis of orthogonal polynomials $\{L^{(\ell)}_{k}(r), k = 0, 1, \ldots\}$, associated with the integration measure $r^{\ell} e^{-r} \rmd r$,
\beq\label{eq:orthogonality}
\int_{0}^{\infty} \rmd r\, r^{\ell} e^{-r} L^{(\ell)}_{k}(r) L^{(\ell)}_{l}(r) = s_{k} \delta_{k, l}\, ,
\eeq
with $\delta_{k, l}$ the Kronecker delta and with $L^{(\ell)}_{k}(r)$ a polynomial of degree $k$ in $r$. The key advantage of this basis is that it allows one to factorize the determinant. Namely, decomposing the monomials $r^{k}$ over it,
\beq\label{eq:alpha_appendix}
r^{k} = \alpha_{k} L_{k}^{(\ell)} + \ldots\, , \qquad \alpha_{k} \neq 0\, ,
\eeq
with the dots standing for sums of polynomials of lower degrees, and using basic properties of the determinant, one may write
\beq
\begin{aligned}
\Phi_{m, n} &\approx \frac{(2|\sigma|)^{mn}}{\C} \det \left[ \alpha_{i-1}\alpha_{j-1} \int_{0}^{\infty} \rmd r\, r^{\ell} e^{-r} L^{(\ell)}_{i-1}(r) L^{(\ell)}_{j-1}(r) \right]_{1\leqslant i, j \leqslant m} = \frac{(2|\sigma|)^{mn}}{\C} \prod_{i=0}^{m-1} \alpha_{i}^2 s_{i} \, .
\end{aligned}
\eeq
In the case at hand, the $L$'s are the generalized Laguerre polynomials,
\beq
L^{(\ell)}_{k}(r) = \frac{r^{-\ell} e^{r}}{k!} \frac{d^{k}}{dr^{k}} (e^{-r} r^{k+\ell}) = (-1)^{k} \frac{r^{k}}{k!} + \mathcal{O}(r^{k-1})\, ,
\eeq
which obey~\eqref{eq:orthogonality} with $s_{k} = (k+\ell)!/k!$ and~\eqref{eq:alpha_appendix} with $\alpha_{k} = (-1)^{k} k!$. One thus immediately concludes that
\beq
\Phi_{m, n} \approx \frac{(2|\sigma|)^{mn}}{\C} \prod_{i=0}^{m-1} (i! (i+\ell)!) = \frac{(2|\sigma|)^{mn}}{\C} \frac{G(m+1) G(n+1)}{G(n-m+1)}\, ,
\eeq
recalling that $\ell = n-m$ and using $G(z+1) = \prod_{i=0}^{z-1} i!$.

\bibliography{biblio}
\bibliographystyle{JHEP}

\end{document}